\documentclass[authoryear,preprint,11pt]{elsarticle} 
\journal{-} 

\usepackage{aas_macros} 
\usepackage{amsmath}
\usepackage{color}
\usepackage{natbib}
\usepackage{graphicx}
\usepackage{subfigure}
\graphicspath{{fig/}}

\def\s{{\rm\,s}}
\def\m{{\rm\,m}}
\def\cm{{\rm\,cm}}
\def\mum{\,\mu{\rm m}}
\def\gm{{\rm\,g}}
\def\kg{{\rm\,kg}}
\def\km{{\rm\,km}}
\def\yr{\rm\,yr}

\def\H2O{\rm\,H_2O}
\def\Op{\rm\,O^+}
\def\Opp{\rm\,O_2^+}
\def\O2{\rm\,O_2}

\DeclareMathOperator\erf{erf} 
\providecommand{\abs}[1]{\lvert#1\rvert} 
\providecommand{\angles}[1]{\langle#1\rangle} 

\begin{document}

\begin{frontmatter}
\title{Particle dynamics in the central ringlet of Saturn's Encke gap}
\author{Kai-Lung~Sun, J\"urgen~Schmidt, Frank~Spahn}
\address{Karl-Liebknecht-Str. 24/25, 14476 Potsdam Golm}

\begin{abstract}
    A kinky and clumpy ringlet shares orbit with the moon Pan in the center of the 320-km wide Encke gap in Saturn's rings \citep{PorcoBakerEtAl2005}.
    The ringlet is mainly composed of micron-sized particles \citep{Showalter1991,HedmanNicholsonEtAl2011}, implying that these particles may be significantly perturbed by non-gravitational forces, which can limit their lifetimes.
    We establish a kinetic model considering the birth, evolution, and death of dust in the Encke central ringlet allowing to evaluate the ringlet optical depth. First, we investigate the generation of dust by micrometeorite impacts (the `\textsl{impact-ejecta}' process) on putative, yet undetected embedded moonlets. Taking into account the orbital evolution under the influence of the relevant perturbation forces, the dominant loss mechanisms are collisions with ring particles in the gap edges, the putative moonlets in the gap, or erosion by sputtering in Saturn's plasma environment.
    However, our results show that this impact-ejecta process alone can only sustain a ringlet of optical depth 3--4 orders of magnitude smaller than the observed values.
    Consequently, other processes must be taken into account. For example, mutual collisions among putative moonlets should produce dust at about the same rate, while further disruption of ejecta, as proposed by \citet{Dikarev1999}, should increase the total amount of particles.
    Furthermore, observations show an azimuthal asymmetry of the material in the Encke gap ringlets (\citealt{FerrariBrahic1997}; M. Srem{\v c}evi{\'c} 2011, private communication). We investigate the scenario proposed by \citet{Hedman2013} that for the Encke central ringlet, the observed asymmetry is mainly due to the combined action of plasma drag and Pan's gravity causing a focusing of dust in the region leading Pan's orbit \citep{Hedman2013}.
\end{abstract}

\end{frontmatter}   


\section{Introduction}
\begin{figure}
        \begin{center}
            \resizebox{0.65\textwidth}{!}{\includegraphics{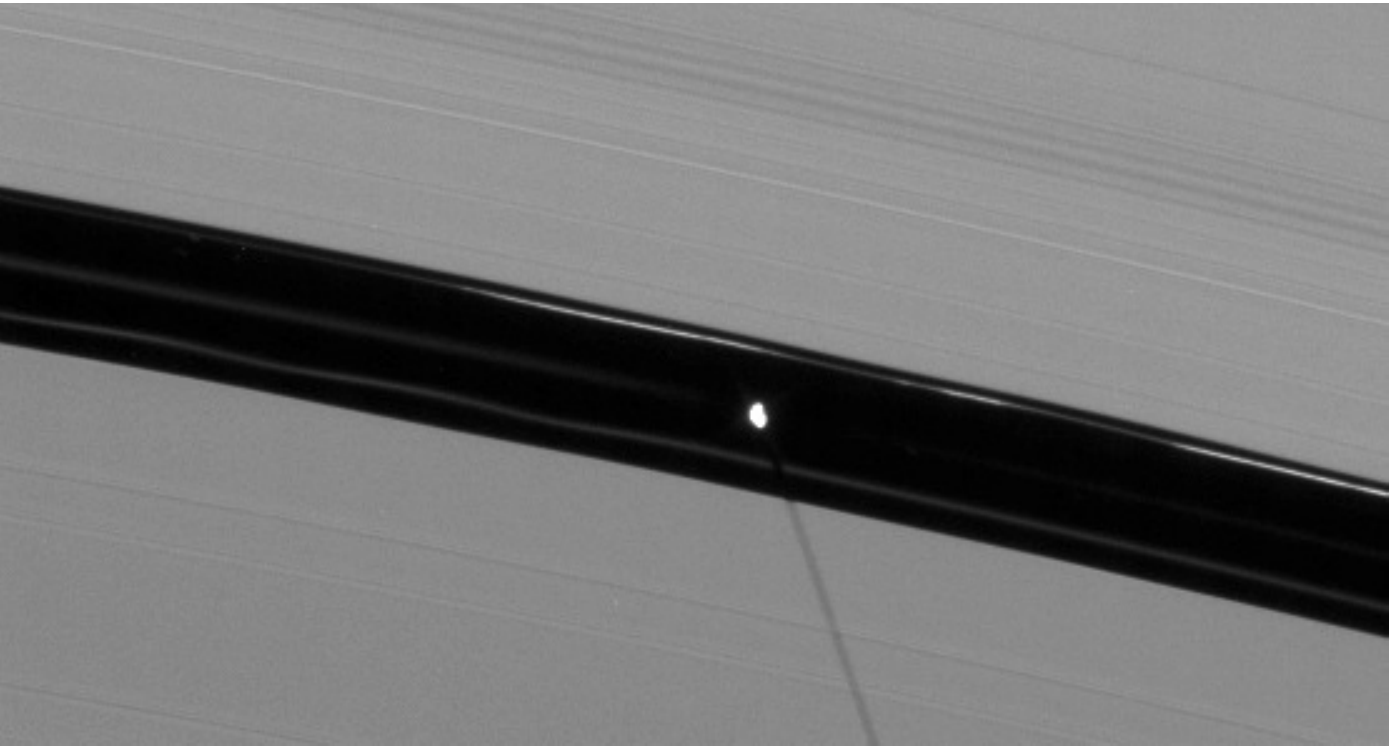}}
            \caption{Pan in the Encke gap (here casting a long shadow over the A ring, when imaged near Saturn's equinox in 2009).
            The bottom (upper) ringlet is the inner (outer) ringlet and the central ringlet shares the orbit of Pan.
            The diffusive ringlet between central and outer ringlet is not visible here.
            One can see the brightness variations in the central and outer ringlet.
            (source: http://photojournal.jpl.nasa.gov/catalog/PIA11581)
            }
            \label{fig:Enckegap}
        \end{center}
    \end{figure}

The Encke gap is a 320~km wide division in Saturn's A ring centered at a planetary distance of 133,581~km. It is maintained by gravitational action of the embedded moon Pan, which shape is approximated by an ellipsoid with axes $17.4\times15.8\times10.4$~km \citep{Showalter1991,PorcoBakerEtAl2005,PorcoThomasEtAl2007}.
    In total, there are four ringlets found in this division, three of them are visible in Fig~\ref{fig:Enckegap}.
    The central ringlet shares it's orbit with Pan.
    There are two further ringlets fainter than the central ringlet, each about 100~km away from the center towards the inner and outer gap edge, which we call the inner and outer ringlets.
    Another more diffusive ringlet is located between central and outer ringlet.

    The optical depths of central, inner, and outer Encke ringlets vary significantly with azimuthal longitude.
    For instance, the clumps in the Encke central ringlet are found to be concentrated in a region 0--60$^{\circ}$ ahead the orbit of Pan with slant optical depth $>0.1$\footnote{It has been suggested that the concept of \textit{normal} optical depth may not be valid for compact structures like the clumps in the F ring and Encke ringlets \citep{HedmanNicholsonEtAl2011}} (\citealt{PorcoBakerEtAl2005}, \citealt{FerrariBrahic1997}; M. Srem{\v c}evi{\'c} 2011, private communication; \citealt{HedmanNicholsonEtAl2011}), while other parts of the ringlet are much fainter.
    \citet{Hedman2013} suggested that the concentration results from a combination of plasma drag and the particle dynamics in the horseshoe region of Pan.

    Observations show strong forward light scattering, indicating that Encke ringlets are composed of micron-sized particles \citep{ShowalterPollackEtAl1992,HoranyiBurnsEtAl2009}. Furthermore, based upon the strength of spectra feature, the so called `Christiansen effect', \citet{HedmanNicholsonEtAl2011} suggested that the fraction of particles with radius smaller than 10$\mum$ in Encke ringlets is less than in F ring or the ringlet in Laplace gap.
    Later in Section~\ref{sec:F_rad} we show that the large eccentricities of small (radii less than 2.4$\mum$) particles induced by solar radiation pressure can explain the lack of few micron particles.

    Apart from the few-micron particles, the lifetimes of larger micron-sized particles (radii between 2.4$\mum$ and 100$\mum$) are restricted by several processes. We will demonstrate that the lifetimes are about 30 years with putative embedded moonlets or about $10^4$ years (for 10$\mum$ radius particles) due to plasma sputtering.
    Consequently, mechanisms must be at work to replenish these narrow dusty ringlets.

    Our main purpose is to understand the origin and evolution of particles in the central ringlet. To this end we explore the hypothesis that Pan, or putative, unseen moonlets in the horseshoe region of Pan, act as dust sources of Encke central ringlet. Such a population of moonlets with size between tens to hundreds of meters would be below the detection limits of the Cassini Imaging Science Subsystem (ISS; the resolution is $\sim$1~km/pixel on the rings, \citealt{PorcoWestEtAl2004}).  Dust is released from Pan and these moonlets through micrometeorite impacts (the impact-ejecta process) and/or in mutual collisions between the moonlets. These processes have been widely invoked to explain the dust formation in other contexts.
    For example, impact-ejecta process is used to explain the dust cloud around Galilean moons \citep{KrivovSremvEtAl2003,SremvKrivovEtAl2005}, Saturn's F ring \citep{ShowalterPollackEtAl1992} and G ring arcs \citep{HedmanBurnsEtAl2007}, while impact-ejecta together with moonlet mutual collisions is proposed to maintain rings of Uranus and Neptune \citep{ColwellEsposito1990grl,ColwellEsposito1990ica}.

    Non-gravitational forces must be taken into account in the dynamics of micron-sized dust.
    Important perturbations for a circumplanetary particle are solar radiation, planetary oblateness, and electromagnetic forces \citep{HoranyiBurnsEtAl1992,Hamilton1993,HedmanBurtEtAl2010}.
    Furthermore, plasma drag should also play a role. An ionosphere mainly composed of $\Op$ and $\Opp$ has been discovered around Saturn's ring system \citep{TokarJohnsonEtAl2005,CoatesMcAndrewsEtAl2005,WaiteCravensEtAl2005}.
    The interaction between ions and dust increase the semi-major axes of dust, probably explaining the observed semi-major axis distribution in Encke central ringlet at certain range of co-rotating longitudes \citep[][Fig. 21]{Hedman2013}.

    In order to describe the generation, evolution, and sustainment of the Encke central ringlet, we establish a kinetic model considering the `birth', evolution, and `death' of dust in the Encke central ringlet.
    The structure of this paper is as follows.
    The kinetic equations are developed in Section~\ref{sec:kinetics}, with short summaries in Section~\ref{sec:sum_dyn} and \ref{sec:sum_kin}.
    In Section~\ref{sec:results}, the results and discussions are presented.
    Conclusions and outlook are given in Section~\ref{sec:conclusion}.

\section{Kinetic model}
    \label{sec:kinetics}

\subsection{Ringlet at quasi-steady state}
    \label{sec:steady_state}
    Assuming for simplicity, a narrow dusty ring with a homogeneous spatial distribution of the dust grain in a well defined region of the ringlet.
    Then, the time evolution of the total number of particles $N$ in this ringlet is given by the kinetic balance
    \begin{equation}
        \label{eq:gl} 
        \frac{dN}{dt} = \dot{N}^{+} - \dot{N}^{-}
    \end{equation}
    where $\dot{N}^{+}$ and $\dot{N}^{-}$ denote the gain and loss rates caused by sources and sinks, respectively.
    Assuming the ringlet reaches a quasi-steady state, $\dot{N}^{+} \approx \dot{N}^{-}$ and the total number of ringlet particles can be estimated as
    \begin{equation}
        \label{eq:steady_state}
        N = \dot{N}^{+} \angles{T_{life}}
    \end{equation}
    where $\angles{T_{life}}$ is the average particle lifetime dominated by the particle sinks and the dynamical life of particles (i.e., they may escape the ringlet).
    This equation can also be written in terms of ringlet geometric optical depth
    \begin{equation}
        \label{eq:steady_state_tau}
        \tau = \dot{\tau}^{+} \angles{T_{life}}
    \end{equation}
    with the optical depth increase rate
    \begin{equation}
        \label{eq:tau_plus}
        \dot{\tau}^{+} = \frac{\dot{N}^{+} \pi \angles{s}^2 }{A_{ring}}
    \end{equation}
    where $\angles{s}$ is the averaged radius of the dust grains and $A_{ring} = 2 \pi a w$ is the total area of the ringlet.
    The value of $A_{ring} \approx 3.2 \times 10^{13} \m^2$ is obtained by setting the semi-major axis to $a = 133,584 \km$ and assuming a ringlet $w = 38 \km$, the latter is equal to the diameter of Pan's Hill sphere. 

    In other words, for an estimate of the number of particles of a ringlet at steady-state, two crucial parameters must be quantified: the source rate $\dot{N}^+$ and the mean particle lifetime $\angles{T_{life}}$, characterizing the particle sources and sinks.
    A similar model was established by \citet{BodrovaSchmidtEtAl2012} for dense rings, where a kinetic balance between adhesion and collisional release of regolith like particles from the surface of parent bodies was considered. The model was used to explain the lower cut-off of particle size Saturn's dense rings.
    Here we follow a similar concept, considering a putative population of skyscraper-sized moonlets in the horseshoe region of Pan acting as major dust source.

\subsection{Constraints on an embedded moonlet population and our assumption}
    \label{sec:assumption}
    An embedded moonlet belt in the Encke central ringlet would act both as source and sink of dust.
    These moonlets, if they exist, would be gravitationally bound to the horseshoe or tadpole region of Pan.
    Their size must be in the range of tens to hundreds of meters in diameter, with upper limit smaller than the detection limit of Cassini Imaging Science Subsystem \citep{PorcoBakerEtAl2005}.
    As suggested by \citet{MurrayBeurleEtAl2008}, most structures in Saturn's F ring are related to Prometheus, Pandora, and a population of embedded moonlets.
    The kinky and clumpy structures of the Encke gap ringlets resemble the F ring structures, and thus may be in a similar result from the gravitational action of such embedded moonlets.

    What is a reasonable guess of the population of these embedded sub-km moonlets? \citet{HedmanNicholsonEtAl2011} have about 87 Cassini VIMS occultation profiles of the Encke ringlets, 24 of them have sufficient  signal-to-noise ratio to assure the detection of the low optical depth ringlet. No (infrared) opaque objects have been identified in occultations.
    The typical Fresnel zone for such an occultation is about 40 meters \citep{BrownBainesEtAl2004}. Therefore the total scanned area on the central ringlet is roughly $40 \times 87 \times w \m^2$, where $w \approx 38 \km$ is width of ringlet. This is about $4.15 \times 10^{-6}$ of the total ringlet area (the $A_{ring}$ mentioned above). Imagine there are 100/10,000 occultation-detectable moonlets embedded in Encke central ringlet, this means the chance to find at least one moonlet in these 87 occultations is about $4.15 \times 10^{-4}$/$4.06\times 10^{-2}$. Therefore it is safe to say that there could have hundreds to thousands of sub-km moonlets.

    Our assumption is: there are 100 moonlets with radius of 250 meters. This makes the total moonlet optical depth $\tau_m \approx 10^{-6}$. Although such assumption seemed to be over-simplified, as we shown in the following section, all we need is the cross section which acts as source and sink of dust for impact-ejecta process. This $\tau_m \approx 10^{-6}$ can also spread into a moonlet population with some steeper size distribution, this does not change the result of impact-ejecta process. Next question is why we choose $\tau_m  \approx 10^{-6}$? In fact, for impact-ejecta process, the source rate and sink rate are both proportional to the total moonlets cross section. So, as long as one don't choose an extreme $\tau_m$ which makes other sources or sinks become important, the ringlet will reach same balance.

\subsection{Sources}
    \label{sec:src}

\subsubsection{Impact-ejecta process}
    \label{sec:source}
\begin{figure}
        \begin{center}
            \resizebox{0.75\textwidth}{!}{\includegraphics{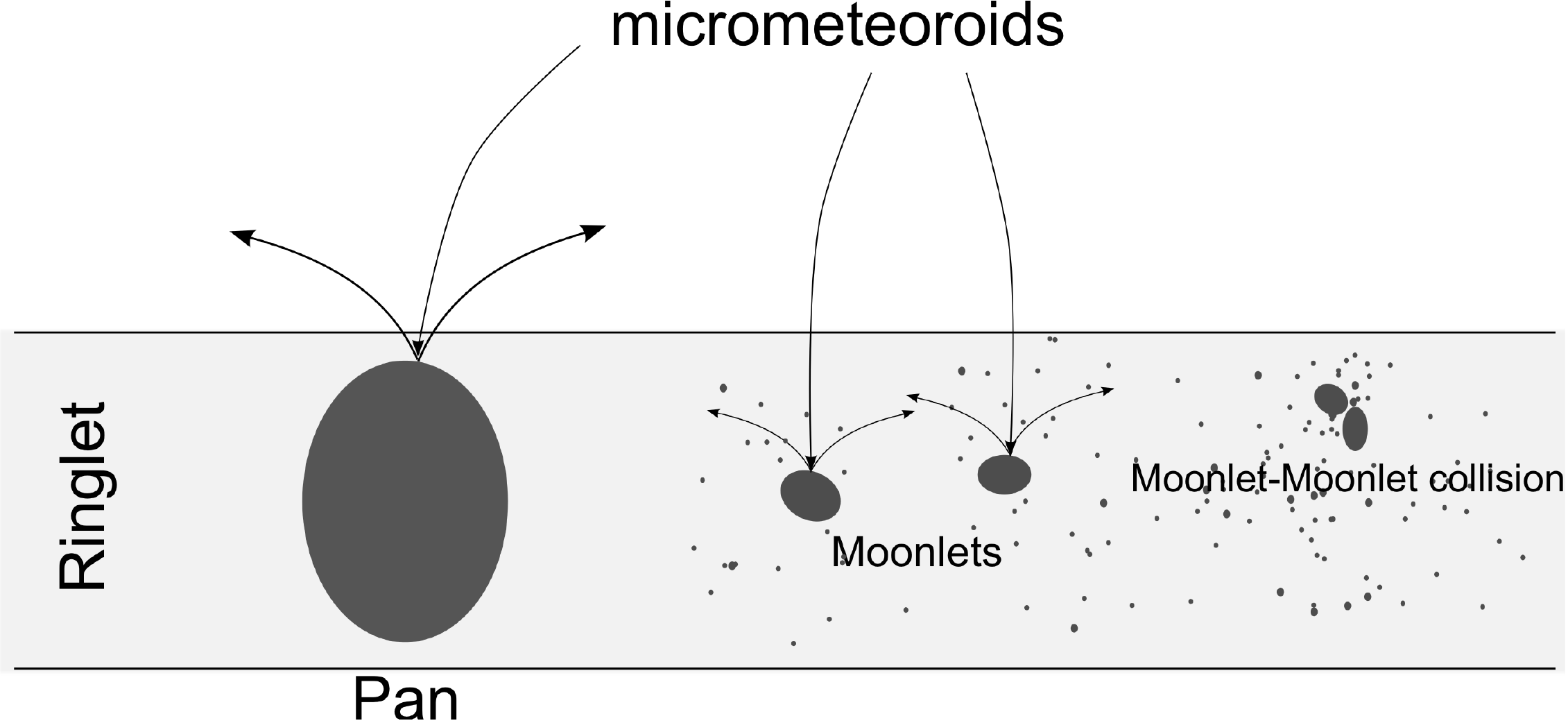}}
            \caption{Schematic figure of particle sources in the central Encke ringlet with putative embedded moonlets. The sources are: impact-ejecta from Pan, impact-ejecta from putative embedded moonlets, and moonlet mutual collisions.
             }
            \label{fig:sources}
        \end{center}
    \end{figure}

    Schematically, the release of dust through the impact-ejecta process from Pan or putative embedded moonlets in the Encke central ringlet is shown in Fig.~\ref{fig:sources}.
    The mass production rate \citep{ShowalterPollackEtAl1992,KrivovSremvEtAl2003,SpahnAlbersEtAl2006} can be parameterized in the form
    \begin{equation}
        \label{eq:mpr} 
        M^+ = F_{imp} Y S\,.
    \end{equation}
    Here, $F_{imp}$ is the mass flux of projectiles dominated by interplanetary dust particles (IDPs), we adopt a fiducial value of $F_{imp} = 1.16 \times 10^{-16} \gm\cm^{-2}\s^{-1}$ (explain later);
    the ejecta yield $Y \approx 26,820$ is the ratio of ejected mass to the mass of the projectile,
    and $S = \pi R_{target}^2$ is the cross section of the targets.
    The yield is obtained base on the experiments and assume a pure icy surface \citep{KrivovSremvEtAl2003,KoschnyGrun2001}. Note that the both $F_{imp}$ and yield are highly uncertain. For example, if there are 10\% of silicate on the surface of target moons, the yield decrease to about 2/3 of this value.

    At Saturn's heliocentric distance, the mass flux of IDPs ($F^{\infty}_{imp}$) is not yet well determined by observations. The infinity sign ($\infty$) indicates this flux is far from Saturn and is not influenced by Saturn's gravity, while the $F_{imp}$ in Eq.~\eqref{eq:mpr} is the flux enhanced by gravitational focusing of Saturn. The gravitational focusing is a function of radial distance to Saturn \citep{SpahnAlbersEtAl2006}.
    Since the flux is not yet well determined, we simply use $F^{\infty}_{imp} = 1.8 \times 10^{-17} \gm\cm^{-2}\s^{-1}$ as fiducial number, same as previous studies \citep{SpahnAlbersEtAl2006,KrivovSremvEtAl2003} which is based on model by \citet{Divine1993}. After consider the gravitational focusing \citep{SpahnAlbersEtAl2006}, we obtain $F_{imp} = 1.16 \times 10^{-16} \gm\cm^{-2}\s^{-1}$ in the orbital range of Encke gap. Note that this impactor flux is likely overestimated and has order(s) of magnitude uncertainties.
    The reason is that, base on the impactor flux above, there should have dust cloud near Rhea. But optical observations by Cassini ISS \citep{TiscarenoBurnsEtAl2010a} and measurements of Cosmic Dust Analyzer (CDA) onboard Cassini \citep{Mochrie2012} both cannot detect dust cloud or ring near Rhea. It has been suggested by \citet{Mochrie2012} that the impactor flux is not stationary, but is varying with time. For example, $F_{imp}$ should be increased while the Saturn system is passing through meteor swarms, same mechanisms for the meteor showers observed on Earth regularly.

    If the ejecta size-distribution is approximated by a power law, then the total production rate of ejecta larger than radius $s$ is given by \citep{SpahnAlbersEtAl2006}
    \begin{equation}
        \label{eq:npr} 
        \dot{N}^{+}_{eje}(>s) = \left( \frac{3-\gamma}{\gamma} \right)
                    \left( \frac{M^{+}}{m_{max}} \right)
                    \left( \frac{s_{max}}{s} \right)^{\gamma}
    \end{equation}
    where $s_{max}$ and $m_{max}$ are the maximum radius and mass of the ejecta. Following again the work by \citet{SpahnAlbersEtAl2006} we take $s_{max} = 100\mum$ and $m_{max} \approx 10^{-5} \kg$ with density $\sim 2400 \kg\m^{-3}$, a size that is believed to represent the peak mass flux of the projectiles, and $\gamma=2.4$ denotes the slope of the cumulative size distribution.

    By assuming the distribution of particle size between $s$ and $s_{max}$ are same in the ringlet as in the ejecta, the effective radius of particle $\angles{s}$ which weighted by the average cross section is
    \begin{equation}
        \label{eq:s_avg}
        \angles{s} =
            \left[ \left( \frac{-\gamma}{2-\gamma}\right) \left( \frac{s_{max}^{2-\gamma} -
                    s_{min}^{2-\gamma}}{s_{max}^{-\gamma} - s_{min}^{-\gamma}}\right)
            \right]^{1/2} \,, \gamma \neq 2\\
    \end{equation}
    where $s_{min}$ is the minimum particle radius.  In the Encke gap, particles smaller than 2.4$\mum$ have very short lifetimes so that they practically do not contribute to the particle population in the central Encke ringlet.  This is because they rapidly develop large eccentricities by the action of solar radiation pressure and are lost to the A ring (see Section~\ref{sec:F_rad} for details). Therefore we use $s_{min} = 2.4 \mum$, $s_{max}=100\mum$, and $\gamma=2.4$ in equation (\ref{eq:s_avg}), which results in $\angles{s} \approx 5 \mum$.

    Not all of the freshly produced ejecta contribute to the ringlet because their initial speed may not be sufficient to overcome the gravity of their parent body. Especially for Pan a fraction of ejecta simply fall back to the satellite. To quantify this effect we describe the initial speed distribution by a power law \citep{KrivovSremvEtAl2003,SpahnAlbersEtAl2006}
    \begin{equation}
        \label{eq:u_distri}
        f_v(v) = \frac{\zeta - 1}{v_0} \left( \frac{v}{v_0} \right)^{-\zeta} H(v-v_0)
    \end{equation}
    with the normalization
    \begin{equation}
        \int_{0}^{\infty} f_v(v) dv = 1
    \end{equation}
    where $H(x)$ is the Heaviside step function and $v_0$ the minimum ejecta speed. Plausible values for the slope of the speed distribution $\zeta$ should lie between 2 and 3. Here $\zeta \approx 2$ would be expected for regolith targets and a larger number $\zeta \approx 3$ is for solid surfaces. The minimum ejecta speed $v_0$ is constrained by the fraction of the projectile's kinetic energy $K_i$ that is converted into the ejecta's kinetic energy $K_e$. We use the relationship between the ratio $K_e/K_i$, yields $Y$, minimum ejecta speed $v_0$ and slope $\zeta$ given by \citep{KrugerKrivovEtAl2000,KrivovSremvEtAl2003}
    \begin{equation}
        \label{eq:KeKi}
        \frac{K_e}{K_i} = Y \left( \frac{\zeta-1}{3-\zeta} \right)
                          \left( \frac{v_0}{v_{imp}} \right)^2
                          \left[ \left( \frac{v_0}{v_{max}} \right)^{\zeta-3} -1 \right]
                          \quad \text{for } \zeta \neq 3
    \end{equation}
    or
    \begin{equation}
        \label{eq:KeKi2}
        \frac{K_e}{K_i} = 2Y \left( \frac{v_0}{v_{imp}} \right)^2
                          \ln \frac{v_{max}}{v_0}
                          \quad \text{for } \zeta=3
    \end{equation}
    where we assume a maximum ejecta speed $v_{max} = 3 \km/\s$ \citep{KrivovSremvEtAl2003}.
    In experiments \citep{Hartmann1985, Asada1985} the ratio $K_e/K_i$ is found in the range between a few percent to several tens of percent. It is larger when the projectile speed is high or the target surface is hard \citep{KrivovSremvEtAl2003}.
    In the Encke gap, the impactor speed $v_{imp}$ is relatively high ($\sim 20 \km \s^{-1}$, due to gravitational focusing of Saturn) and the target surfaces are probably regolith-like. Hence we adopt a moderate value $K_e/K_i=0.1$.
    Using this value together with $\zeta=2$ (for regolith surfaces) and $Y \approx 26,820$, we obtain $v_0 \approx 1\m \s^{-1}$ from Eq.~\eqref{eq:KeKi}.
    Because of the power law distribution of initial speeds, most of the particles are ejected with speeds only marginally larger than $v_0$.

    The ejecta speed of 1$\m\s^{-1}$ is larger than the escape velocity of sub-km moonlets, but smaller than that of Pan. Three body effects can be incorporated in an approximate manner, resulting in the expression for the escape velocity \citep{SpahnAlbersEtAl2006} 
    \begin{equation}
        v_{esc} = \sqrt{2GM_{moon} \left( \frac{1}{R_{moon}} - \frac{1}{R_{Hill}}
                  \right)}
        \label{eq:vesc}
    \end{equation}
    where $R_{moon}$ and $M_{moon}$ are the radius and mass of the moon, $R_{Hill} = r \sqrt[3]{M/(3M_s+M)}$ is the Hill radius of the moon and $r$ the radial distance to Saturn. The moonlet mass is calculated by assuming a moonlet density of 410 kg/m$^3$, as was estimated for Pan \citep{PorcoThomasEtAl2007}. Therefore, for Pan and a 250-meter-radius moonlet in the Encke gap, the escape velocities are about 3.6~m/s and 0.06~m/s, respectively.

    The fraction of escaping ejecta is related to the escape velocity $v_{esc}$ and the initial speed distribution of ejecta \citep{SpahnAlbersEtAl2006}
    \begin{equation}
        \label{eq:escape_frac}
        \dot{N}^+_{esc}/\dot{N}^+_{eje} = \left\{
            \begin{array}{l l}
                (v_0/v_{esc})^{\zeta-1} & \quad \text{for } v_0 < v_{esc} \\
                1                       & \quad \text{for } v_0 \geq v_{esc}.
            \end{array}
            \right.
    \end{equation}
    Therefore, only some ($\sim$28\%) ejecta from Pan can escape, while all ejecta from 250 meter moonlets can escape. 
    For this simple estimate we ignored the non-spherical shape of Pan and the detailed three body dynamics in the Hill sphere, which cause the escape velocity to vary with the launching position of ejecta \citep{BanaszkiewiczIp1991}.

    As shown above, dust produced by impact-ejecta cannot all escape Pan.
    Furthermore, for those escape Pan, it is still unlikely that they can contribute to the Encke central ringlet because they are gravitationally scattered by Pan.
    We use $\eta$ to represent the fraction of escaped ejecta from Pan that enter the ringlet, which should a be very small number and is quantified to be 0.125\% in Section~\ref{sec:res_Pan}.
    On the other hand, ejecta escape from smaller moonlets are already in horseshoe orbit of Pan and the gravity of these less massive moonlets is not important, therefore we assume all ejecta from moonlets contribute to the ringlet.

    Putting all the parameters into Equations~\eqref{eq:tau_plus}--\eqref{eq:escape_frac}, we obtain the rates for dust production $\dot{N}^+_{eje}$, dust escape $\dot{N}^+_{esc}$, and optical depth increase $\dot{\tau}^+$, and all summarized in Table~\ref{tbl:n_plus}. Here we used $S \approx$ 612 km$^2$ for the cross section of Pan and $S \approx$ 20 km$^2$ for the moonlet belt. Since the value of $\eta$ must be very small and we do not see another mechanisms to transport these scattered particles back to the ringlet, we conclude that Pan cannot be the major source of Encke central ringlet.
    \begin{table}
        \centering
        \begin{tabular}{ccccc}
            \hline
            Source   & $S$ ($km^2$) & $\dot{N}^+_{eje}$ ($s^{-1}$) & $\dot{N}^+_{esc}$ ($s^{-1}$) & $\dot{\tau}^+$ ($\yr^{-1}$) \\
            \hline
            Pan      & 612 & $3.0 \times 10^9$ & $8.3 \times 10^8$ & $2.9\eta \times 10^{-7}$ \\
            Moonlets & 20  & $9.6 \times 10^7$ & $9.6 \times 10^7$ & $3.4 \times 10^{-8}$ \\
            \hline
        \end{tabular}
        \caption{Dust production rates due to the impact-ejecta process in the Encke central ringlet.
            The putative moonlets have a total cross section $S$ of 20$\km^2$, or optical depth of $10^{-6}$.
            $\dot{N}^+_{eje}$ is the dust production rate; $\dot{N}^+_{esc}$ is the rate that dust escaping the source moon; $\dot{\tau}^+$ is the optical depth increase rate.
            The factor $\eta$ describes the fraction of ejecta from Pan that can contribute to the ringlet (i.e., particles remaining in horseshoe region), it's value is roughly estimated to be about 0.125\% (see Section \ref{sec:res_Pan}).
        }
        \label{tbl:n_plus}
    \end{table}

A major disadvantage is that the parameters of the impact-ejecta process are poorly known.
The major uncertainties are the projectile flux $F_{imp}$ and the yield $Y$ \citep{KrivovSremvEtAl2003}, which may be uncertain by orders of magnitude. The ejecta size and velocity distributions are also poorly constrained, adding to the uncertainty in the distribution of freshly produced dust.

\subsubsection{Secondary debris: Collisional disruption of ejecta}
    \label{sec:second}
    The ejecta produced in hypervelocity impacts can serve themselves as secondary dust sources when they are fragmented in mutual collisions and micrometeoroid impacts. Such processes, in general, do not add mass to the system but tend to increase the total cross section. The mechanism was suggested by \citet{DikarevKrivovEtAl2006} to enhance the dust production rate of the impact-ejecta process of certain Jovian and Saturnian moons. In the following we briefly outline their model.

    After disruption, the mass distribution of particle fragments is assumed to be similar to the distribution of impact-ejecta (Eq.~\eqref{eq:npr})
    \begin{equation}
        \label{eq:npr1} 
        N_1(m_1; >m) = \frac{1-\alpha}{\alpha \beta} \left( \frac{\beta m_1}{m} \right) ^{\alpha}
    \end{equation}
    where $m_1$ is the mass of the disrupted particle, $\alpha = \gamma / 3$ is the slope of the mass distribution and $\beta$ is the mass ratio of the biggest fragment and the parent particle. A value of $\beta \sim 10^{-3}$ is plausible for an impact speed on the order km~s$^{-1}$ \citep{KrivovSremvEtAl2005}.

    The inter-particle collision rate is estimated by the rate of particles re-accreted to the moon $N(>m)T_{acc}^{-1}$ and then scale it by $\varepsilon A_b / S$.
    The re-accretion time $T_{acc}$ is obtained from the belt model developed by \citet{KholshevnikovKrivovEtAl1993}, $A_b/S$ is the ratio of the cumulative cross section of the belt particles to the cross-section of the moon.
    The variation of $A_b$ due to the disruption of ejecta is not considered in \citet{KholshevnikovKrivovEtAl1993} belt model, therefore an additional factor $\varepsilon$ is included
    (see \citealt{DikarevKrivovEtAl2006} for more discussion in how to choose $\varepsilon$).
    Hence, the dust production rate caused by mutual collisions between ejecta from the moons reads
    \begin{equation}
        \label{eq:intra}
        \dot{N}_{int}(>m_*) = - \int_{m_{min}}^{m_{max}} N_1(m; m>m_*) \frac{\varepsilon A_b}{S} dN(>m) T_{acc}^{-1}.
    \end{equation}
    The enhancement factor caused by this process is
    \begin{equation}
        \label{eq:Dint}
        D_{int} \equiv \frac{\dot{N}_{int}}{\dot{N^+}}.
    \end{equation}

    In the case of Encke central ringlet, we adopt $\beta = 10^{-3}$, a minimum size $s_{min} = 2.4 \mum$, and the total cross section of moonlet belt $S$ from Table~\ref{tbl:n_plus} to obtain the $\dot{N}_{int}$.
    With the $\dot{N}_{int}$ and the $\dot{N}_{esc}^+$ of moonlet belt in Table~\ref{tbl:n_plus}, we find the enhancement factor of the secondary source is $D_{int} \approx 9.7$.
    Note that this process also provides extra target surface for the impact-ejecta process itself, which is also given in \citet{DikarevKrivovEtAl2006}. But in our case the enhancement factor is much smaller ($\sim$0.03) than the one for inter-particle collisions and therefore it is ignored here.
    In conclusion, by applying Eq.~\eqref{eq:npr1} - \eqref{eq:Dint}, the secondary debris model of \citet{DikarevKrivovEtAl2006}, the disruption of ejecta can increase the total cross section or optical depth of the Encke central ringlets by about 10 times.

\subsubsection{Mutual collisions between moonlets}
    \label{sec:collision}
    Apart from impact-ejecta process and the enhancement discussed in the previous section, mutual collisions between the parent bodies can also deliberate dust. For instance, dust from regolith layers covering the parent bodies can be released during moonlet collisions \citep{BodrovaSchmidtEtAl2012,ColwellEsposito1990grl,ColwellEsposito1990ica}.

    The dust production rate caused by low-velocity collision between moonlets can be written as
    \begin{equation}
        \label{eq:mc}
        \dot{N}^+_{mc} = \nu N_{col}
    \end{equation}
    where $\nu$ is the collision frequency between moonlets and $N_{col}$ is the amount of dust released per collision. For small number of moonlets (small moonlet optical depth $\tau_m$) the former can be roughly approximated by $\nu \approx 3\tau_{m}\Omega_k$ with $\Omega_k$ the angular velocities of these moonlets, this suggests the moonlet collisional timescales $ 1/\nu \sim 80$ years for $\tau_{m} = 10^{-6}$.
    A crude estimate of the dust production rate $N_{col}$ in a low-velocity collision between two moonlets is that about 12\% of the regolith layer is released in one such collision \citep{CanupEsposito1995}. Here we assume that these moonlets in Encke central ringlet are underdense aggregates, nearly filling their Hill sphere and that the regolith layer occupies 50\% of the moonlet radius.

    The simplified assumption of a moonlet belt with only one size is not applicable here.
    This is because for a fixed total moonlet cross section, their total volume, and thus the regolith layer, vary significantly for a different size distribution. For the sake of definiteness we use a plausible size distribution $n(s) \propto s^{-\gamma}$ with $\gamma=3$ and radii ranging from $s=$10--250~m, with a fixed optical depth $\tau_m = 10^{-6}$.

    Using these assumptions with Eqs.~\eqref{eq:tau_plus} and \eqref{eq:mc}, we obtain for the dust creation in moonlet collisions $\dot{\tau}^{+} \approx 2.6 \times 10^{-8} \yr^{-1}$, which is about the same order of magnitude as the production at the moonlets due to the impact-ejecta process (Table~\ref{tbl:n_plus}).
    However, this estimate is very crude. Further investigation must be undertaken to constrain poorly known parameters, such as the amount of dust resting on moonlets, the fraction of dust release per collision and the moonlet size distribution to derive a more reliable, quantitative estimate.

\subsection{Dynamics}
    \label{sec:dynamics}
    After explaining the generation of ringlet particles, the next step is to consider their orbital evolution.
    For micron-meter sized particles the evolution is dominated by the gravity of Saturn and Pan, but perturbing forces can be crucial as well. For example, a lower cut off for the particle size and some aspects of the observed azimuthal asymmetry in the central ringlet can be explained by perturbations acting on micron-sized dust.

The main perturbing forces near Saturn's main ring are the Saturn's oblateness, solar radiation pressure and Lorentz force \citep{HoranyiBurnsEtAl1992,Hamilton1993,HedmanBurtEtAl2010}. Apart from these, the ionosphere near Saturn's main ring \citep{TokarJohnsonEtAl2005,CoatesMcAndrewsEtAl2005,WaiteCravensEtAl2005} exerts a plasma drag which can play an important role for dust dynamics. With all these forces, the equation of motion for micron-sized dust reads
    \begin{equation}
      \label{eq:motion}
      \begin{split}
        m \ddot{\mathbf{r}} = & -m \nabla{\Phi} -
          G M_p m \frac{\mathbf{r}-\mathbf{r}_{p}}{\abs{\mathbf{r}-\mathbf{r}_{p}}^3} \\
        & + q \left[(\dot{\mathbf{r}} - (\Omega_s \times \mathbf{r})) \times \mathbf{B} \right] -
        \frac{I_{\odot} \sigma Q_{pr}}{c} \hat{e}_\odot + \mathbf{F}_{C} + \mathbf{F}_{D}
      \end{split}
    \end{equation}
    with the particle mass $m$ and its acceleration $\ddot{\mathbf{r}}$.
    On the right hand side, we have the gravity of oblate Saturn $\nabla\Phi$, where the gravitational potential $\Phi$ is given in Eq.~\eqref{eq:potential}.
    The second term is Pan's gravity ($\mathbf{F}_{Pan}$), where $M_p$ and $\mathbf{r}_p$ are mass and position of Pan and $G$ is the gravitational constant.
    The third term is the Lorentz force ($\mathbf{F}_{L}$), caused by the relative motion between the particle and Saturn's magnetic field. Charge of particle is denoted by $q$ and the corotational electric field reads $(\Omega_s \times \mathbf{r}) \times \mathbf{B}$, with the spin rate of Saturn, $\Omega_S$ and the Saturn's magnetic field $\mathbf{B}$.
    Solar radiation pressure ($\mathbf{F}_{\odot}$) is the fourth term,
    where $I_{\odot} \approx 14 \,\mathrm{W} / \mathrm{m}^2$ is the solar energy flux at Saturn,
    $c$ is the speed of light,
    $Q_{pr}$ is the radiation pressure efficiency which is of order of unity, and is about 0.3 for icy particles with radius several micrometer or larger \citep{BurnsLamyEtAl1979}. Further,
    $\sigma$ denotes the cross section of the particle,
    and $\hat{e}_\odot$ is a unit vector pointing from the particle toward the Sun.
    The last two terms are Coulomb drag, $\mathbf{F}_{C}$, exerted by ambient plasma on charged particles and the direct force caused by physical collisions with plasma particles, $\mathbf{F}_{D}$. Both components of the plasma drag are discussed in Section~\ref{sec:plasma}.

    Encke central ringlet particle orbits are dominated by the point mass gravity of Saturn, $\mathbf{F}_s = - G M_s m (\vec{r} - \vec{r}_s)/(\abs{\vec{r}-\vec{r}_s}^3)$,
    and Pan, where $M_s$ is mass of Saturn and $\vec{r}_s$ is location of Saturn. The perturbing forces entering Eq.~\eqref{eq:motion} are small compared with $\mathbf{F}_s$. In this case, the particle motion can be described in terms of perturbed Keplerian orbits with perturbation equations for the time evolution of the elements of the orbit. For small eccentricities and inclinations, the perturbation equations are \citep{Burns1976}
    \begin{equation}
      \label{eq:perturba}
      \frac{da}{dt} = \frac{2}{n} 
                      \left[ F_R e \sin f + F_T(1+e \sin f) \right]
    \end{equation}
    \begin{equation}
      \frac{de}{dt} = \frac{1}{an}  
                      \left[F_R \sin f + F_T ( \cos f + \cos E ) \right]
    \end{equation}
    \begin{equation}
      \frac{di}{dt} = \frac{1}{an}  
                      \left[\frac{F_N \cos (\varpi-\Omega+f)}{(1 + e \cos f)} \right]
    \end{equation}
    \begin{equation}
      \frac{d\Omega}{dt} = \frac{1}{an} 
                           \left[\frac{F_N \sin (\varpi-\Omega+f)}{\sin i ( 1 + e \cos f)} \right]
    \end{equation}
    \begin{equation}
      \label{eq:perturbW}
      \frac{d\varpi}{dt} = \frac{1}{ane} 
                           \left[-F_R \cos f + 2 F_T \sin f \right].
    \end{equation}
    Here, $a$ is semi-major axes, $e$ is eccentricity, $i$ is inclination, $\varpi$ and $\Omega$ are the longitudes of pericenter and ascending node, respectively, $f$ is the true anomaly, $n$ the mean motion, and $E$ the eccentric anomaly. The disturbing force $\mathbf{F} = F_R \mathbf{\hat{e}_R} + F_T \mathbf{\hat{e}_T} + F_N \mathbf{\hat{e}_N}$ is decomposed into a component acting in radial direction ($\mathbf{\hat{e}_R}$), transverse to the radial direction in the orbital plane ($\mathbf{\hat{e}_T}$), and normal to the orbital plane ($\mathbf{\hat{e}_N}$).

\subsubsection{Gravity, Oblateness and Lorentz force}
    \label{sec:grav_lore}
    The higher order terms of gravity and the Lorentz force both affect the precession rates of the orbital nodes and the pericenter \citep{HoranyiBurnsEtAl1992,Hamilton1993}.
    The gravitational potential of Saturn can be expanded in an infinite series as \citep[e.g.,][]{MurrayDermott1999}
    \begin{equation}
      \label{eq:potential}
      \Phi = - \frac{G M_s}{r} \left( 1 + \sum_{j=2}^{\infty} J_i
                              \left( \frac{R_s}{r} \right)^i P_i(\cos{\Theta})
                            \right)
    \end{equation}
    where $M_s$ is the mass of Saturn, $R_s$ is equatorial radius of Saturn, $J_i$ are the zonal harmonic coefficients, and $P_i(\cos\Theta)$ are the Legendre polynomials. Here, $\Theta$ is the latitude of the particle in a spherical coordinate system centered at the planet.
    For Saturn, the odd zonal harmonics are small and the dominant first two even coefficients are $J_2 = 0.01629071$ and $J_4 = -0.00093583$ \citep{JacobsonAntreasianEtAl2006}. Higher order terms are ignored.

    The apsidal precession rate caused by the oblateness of Saturn is approximately \citep{Greenberg1981,HoranyiBurnsEtAl1992}
    \begin{equation}
        \label{eq:precess_J2}
        \begin{split}
            \dot{\varpi}_{J2} = & n \left[ \frac{3}{2} J_2 \left( \frac{R_s}{a} \right)^2
                - \frac{15}{4} J_4 \left( \frac{R_s}{a} \right)^4 \right] \\
            \approx & \left[51.4 \left( \frac{R_s}{a} \right)^{3.5} + 7.2 \left( \frac{R_s}{a} \right)^{5.5} \, \right]\mathrm{deg/day}\,,
        \end{split}
    \end{equation}
while the apsidal precession rate induced by the Lorentz force reads \citep{HoranyiBurnsEtAl1992}
    \begin{equation}
        \label{eq:precess_L}
        \begin{split}
        \dot{\varpi}_{L} = & -2 \frac{qB_0}{m} \left( \frac{R_s}{a} \right)^3 \\
            \approx & 5.5 \left( \frac{R_s}{a} \right)^{3}
                \left( \frac{\varphi}{1V} \right)
                \left( \frac{1 \mum}{s} \right)^2 \, \mathrm{deg/day}\,.
        \end{split}
    \end{equation}
    Here $B_0 = -0.21 $ Gauss is the magnetic field strength at Saturn's equator, and $\varphi$ is the equilibrium surface potential of particle in a given charging environment. Note that for magnetic field, both the derivation of $\dot{\varpi}_L$ and our equation of motion Eq.~\eqref{eq:motion} use the assumption of aligned dipole. This simplified model is enough for our purpose since Lorentz force is relatively not important ($\dot{\varpi}_L < \dot{\varpi}_{J2}$, also shown later in Fig.~\ref{fig:comparison_perturbation}) in our case. An more accurate magnetic field model which uses the zonal spherical harmonic coefficients to describe magnetic fields can be found in \citet{BurtonDoughertyEtAl2009}.

    In the Encke gap one obtains $\dot{\varpi}_{J2} \approx 3.2 \,\mathrm{deg/day}$ and $\dot{\varpi}_{L} \approx -0.02 \,\mathrm{deg/day} $ (for $s=10\mum$ and $\varphi = -4$~V as discussed in Section~\ref{sec:plasma}). The precession rate is clearly dominated by $\dot{\varpi}_{J2}$ unless the charge is much larger than we expect or the particle radius is less than a few microns ($\approx 0.8 \mum$ for $\varphi=-4$~V).

\subsubsection{Solar radiation pressure}
    \label{sec:F_rad}
    Solar radiation pressure have usually strong effect on the evolution of orbital eccentricity and inclination of micron-sized particles. Depending on the rates of nodal and pericenter precession the eccentricities and inclinations may grow and decay periodically. Moreover, for the ensemble of dust particles in a ring the \emph{average} locations of the longitude of pericenter and the ascending node can be locked at a fixed range of subsolar longitudes \citep{HoranyiBurns1991,HoranyiBurnsEtAl1992,HedmanBurtEtAl2010}, giving the ring as a whole an effective eccentricity and inclination. The latter could oscillate significantly in one planetary season. Such a heliotropic behavior has been observed for a dusty ringlet in the Laplace gap in the Cassini Division \citep{HedmanBurtEtAl2010}. Here we follow the derivations in \citet{HedmanBurtEtAl2010} for the orbit averaged time evolution of orbital elements under the effect of solar radiation pressure, including the effect of the planet's shadow, and we apply the model to central ringlet in the Encke gap.

\begin{figure}
        \centering
        \begin{subfigure}
            \centering
            \includegraphics[width=0.49\textwidth]{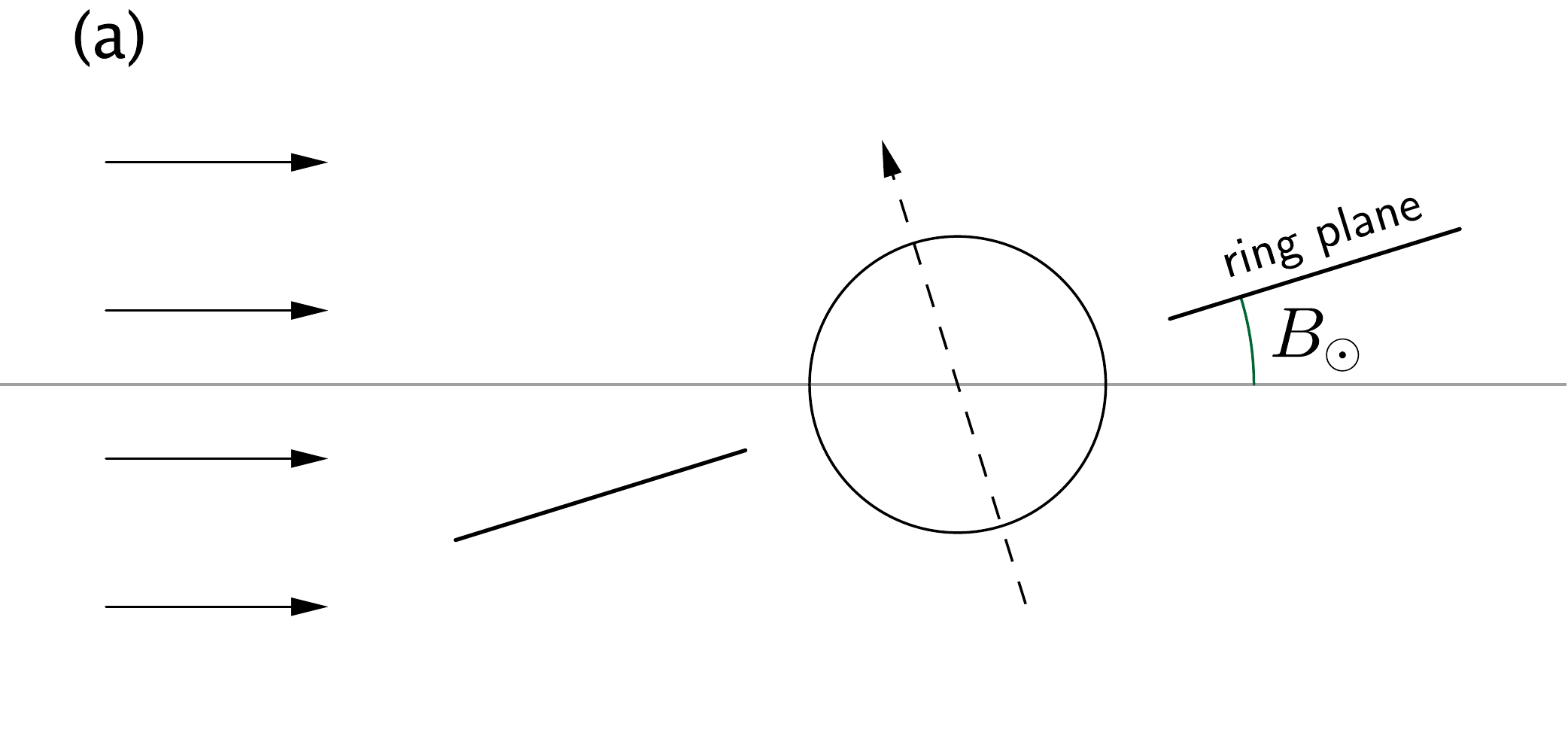}
        \end{subfigure}
        \begin{subfigure}
            \centering
            \includegraphics[width=0.49\textwidth]{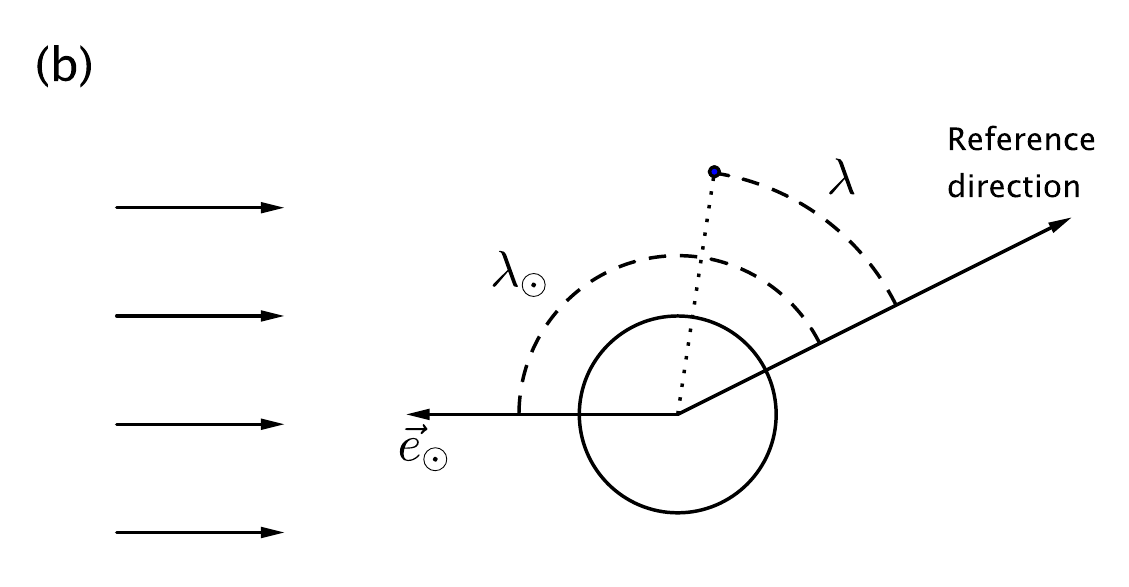}
        \end{subfigure}
        \caption{Solar elevation angle $B_\odot$ and longitude of the Sun $\lambda_\odot$.
            Saturn is represented by the circle and the Sun shines from left.
            \textit{Left panel}: View from side, the gray horizontal line is the ecliptic plane and the dashed arrow denotes Saturn's spin axis.
            \textit{Right}: View from north ecliptic pole; $\lambda$ is the longitude of a particle orbiting Saturn.
        }
        \label{fig:sun}
    \end{figure}

In the equation of motion \eqref{eq:motion}, the normal, radial, and longitudinal components of the solar radiation force $\mathbf{F}_\odot = -(I_0 \sigma Q_{pr} / c) \hat{e}_\odot$  read
    \begin{equation}
        F_N = - F_{\odot} \sin B_\odot
    \end{equation}
    \begin{equation}
        F_R = - F_{\odot} \cos B_\odot \cos(\lambda - \lambda_{\odot})
    \end{equation}
    \begin{equation}
        F_T = + F_{\odot} \cos B_\odot \sin(\lambda - \lambda_{\odot})\,.
    \end{equation}
    Here, $\lambda$ is the longitude of a particle in the ring plane measured from the Saturn center. In this frame the Sun has instantaneous longitude $\lambda_{\odot}$ and elevation angle ${B_\odot}$, the latter varies between $-26.73$ and $+26.73^\circ$ from the ring plane (both shown in Fig.~\ref{fig:sun}).  Expressing the true anomaly as $f = \lambda - \varpi$, the perturbation equations Eq.~\eqref{eq:perturba}-\eqref{eq:perturbW} read
    \begin{equation}
      \label{eq:a}
      \frac{da}{dt} = 2 a n \frac{F_\odot \cos B_\odot}{F_S}
                     \left[
                     e \sin (\varpi - \lambda_\odot) + \sin (\lambda - \lambda_\odot)
                     \right]
    \end{equation}
    \begin{equation}
      \frac{de}{dt} = n \frac{F_\odot \cos B_\odot}{F_S} \left[
                      3 \sin (\varpi - \lambda_\odot) + \sin(2 \lambda - \varpi - \lambda_\odot)
                      \right]
    \end{equation}
    \begin{equation}
      \frac{d\varpi}{dt} = \frac{n}{e} \frac{F_\odot \cos B_\odot}{2 F_S} \left[
                      3 \cos(\varpi - \lambda_\odot) - \cos(2 \lambda - \varpi - \lambda_\odot)
                      \right]
    \end{equation}
    \begin{equation}
      \frac{di}{dt} = - n \frac{F_\odot \sin B_\odot}{F_S} \left[
                      \cos(\lambda - \Omega)
                      \right]
    \end{equation}
    \begin{equation}
      \label{eq:O}
      \frac{d\Omega}{dt} = - \frac{n}{\sin i} \frac{F_\odot \sin B_\odot}{F_S} \left[
                      \sin(\lambda - \Omega)
                      \right] \text{~.}
    \end{equation}

    Next step is to average the evolution equations over one particle orbit, taking into account the shadow of Saturn. The underlying assumption is that  the time evolution takes place on two well separated timescales. The change of $\lambda$, in course of the orbital revolutions takes place on the faster Keplerian timescale. In contrast, the evolution of the orbital elements is taking place on a much slower timescale, owing to the smallness of the perturbation forces. We apply an average of element $X$ over one particle orbit  in the form
    \begin{equation}
      \angles{X} = \frac{1}{2 \pi}
                      \int^{+ \pi (1 - \epsilon)}_{- \pi (1 - \epsilon)}
                      X
                      d(\lambda - \lambda_{\odot})
    \end{equation}
    where the particle is illuminated (by solar radiation) when $\abs{\lambda - \lambda_\odot} < \pi (1 - \epsilon)$. Here, $\epsilon\ge 0$ is the fraction of the particle orbit lying in the planetary shadow. By averaging Eqs.~\eqref{eq:a}-\eqref{eq:O}, we arrive at the evolution equations for the averaged, osculating orbital elements
    \begin{equation}
      \angles{\frac{da}{dt}} = n a e \left [
                               2 d(\epsilon) \frac{F_\odot \cos{B_\odot}}{F_S}
                               \right] \sin (\varpi - \lambda_\odot)
    \end{equation}
    \begin{equation}
      \angles{\frac{de}{dt}} = n \left [
                               \frac{3}{2} f(\epsilon) \frac{F_\odot \cos{B_\odot}}{F_S}
                               \right] \sin (\varpi - \lambda_\odot)
    \end{equation}
    \begin{equation}
      \angles{\frac{d\varpi}{dt}} = \frac{n}{e} \left [
                               \frac{3}{2} f(\epsilon) \frac{F_\odot \cos{B_\odot}}{F_S}
                               \right] \cos (\varpi - \lambda_\odot) + \dot{\varpi}_0
    \end{equation}
    \begin{equation}
      \angles{\frac{di}{dt}} = -n \left [
                               g(\epsilon) \frac{F_\odot \sin{B_\odot}}{F_S}
                               \right] \cos (\Omega - \lambda_\odot)
    \end{equation}
    \begin{equation}
      \angles{\frac{d\Omega}{dt}} = \frac{n}{\sin i} \left [
                               g(\epsilon) \frac{F_\odot \sin{B_\odot}}{F_S}
                               \right] \sin (\Omega - \lambda_\odot) + \dot{\Omega}_0
    \end{equation}
    with the definitions $d(\epsilon) = 1 - \epsilon$, $f(\epsilon) = 1 - \epsilon + \sin(2\pi\epsilon)/6\pi$, and $g(\epsilon) = \sin(\pi\epsilon)/\pi$ \citep{HedmanBurtEtAl2010}. The precession rates $\dot{\varpi}_0$ and $\dot{\Omega}_0$ include the effect of planetary oblateness and the Lorentz force (Eqs.~\eqref{eq:precess_J2} and \eqref{eq:precess_L}).

    This set of equations can be solved under the assumption that the solar elevation angle $B_{\odot}$ and the particle's mean motion $n$ change slowly (the change of $n$ is caused by change of semi-major axes)  compared to the orbital period of ringlet particles.
    The solution can be obtained by transforming the set of equations to the form of harmonic oscillator with new set of variables: $ h = e \cos{(\varpi - \lambda_\odot)}$, $ k = e \sin{(\varpi - \lambda_\odot)}$, $ p = i \cos{(\Omega - \lambda_\odot)}$, $ q = i \sin{(\Omega - \lambda_\odot)}$ \citep{HoranyiBurnsEtAl1992,HedmanBurtEtAl2010}.
    For initial conditions $e(t=0)=0$ and $i(t=0)=0$ the solutions are
    \begin{equation}
        e = \frac{n}{\dot{\varpi}_0'} \left[
              3 f(\epsilon)  \frac{F_{\odot}}{F_S} \cos{B_{\odot}}
            \right]
            \sin{(\dot{\varpi}_0 t / 2)}
        \label{eq:e}
    \end{equation}
    \begin{equation}
        \varpi - \lambda_\odot = \bmod \left(\frac{\dot{\varpi}_0 t}{2}, \pi\right) + \frac{\pi}{2}
        \label{eq:varpi}
    \end{equation}
    \begin{equation}
        i = 2 \frac{n}{\dot{\Omega}_0'} \left[
              g(\epsilon) \frac{F_{\odot}}{F_S} \sin{\abs{B_{\odot}}}
            \right]
            \sin{(\dot{\Omega}_0 t / 2)}
        \label{eq:i}
    \end{equation}
    \begin{equation}
        \Omega - \lambda_\odot = \bmod\left(\frac{\dot{\Omega}_0 t}{2}, \pi\right) + \pi  \frac{B_\odot}{\abs{B_\odot}} \,.
        \label{eq:Omega}
    \end{equation}
    The primed quantities $\dot{\varpi}_0'=\dot{\varpi}_0 - \dot{\lambda}_\odot$ and $\dot{\Omega}_0' = \dot{\Omega}_0 -  \dot{\lambda}_\odot$ include the slow seasonal variation of the solar longitude ($\dot{\lambda}_\odot$). In practise the term $\dot{\lambda}_\odot$ is very small ($\sim 12^\circ/\yr \approx 0.03^\circ$/day) in comparison to the precession rates dominated by planetary oblateness ($\dot{\varpi}_0 \approx \abs{\dot{\Omega}_0} \approx 3.2^\circ$/day), and thus will be ignored for the current problem.

Equations~\eqref{eq:e} and \eqref{eq:i} indicate that $e$ and $i$ are oscillating between zero and a maximum absolute value. The other two equations tell that $\varpi$ and $\Omega$ are locked at certain longitudes relative to that of the Sun, a behavior that has been dubbed `heliotropic' by \citet{HedmanBurtEtAl2010}.
In the Encke gap $n \approx 626^\circ$/day and $\dot{\varpi}_0' \approx \abs{\dot{\Omega}_0'} \approx 3.2^\circ$/day.  Hence the maximum values of eccentricity  and inclination are
    \begin{equation}
        \label{eq:forced_e_2}
        e_{max} \approx 9.7 \times 10^{-3} f(\epsilon) \cos(B_{\odot}) \frac{Q_{pr}}{s/1\mum{}}
    \end{equation}
    \begin{equation}
        \label{eq:forced_i_2}
        i_{max} \approx 6.5 \times 10^{-3} g(\epsilon) \sin(\abs{B_{\odot}}) \frac{Q_{pr}}{s/1\mum{}}\,.
    \end{equation}
    The shadow-fraction $\epsilon$ varies between 15\% (when $B_\odot = 0^\circ$) and 4\% (when $B_\odot \approx 27^\circ$) in the Encke gap. Therefore, $f(\epsilon) \cos (B_{\odot})$ ranges from 0.87 to 0.90, whereas $g(\epsilon) \sin (\abs{B_{\odot}})$ spans values between 0 and 0.035. In other words, over a Saturnian year, the maximal eccentricity is almost constant while the maximal inclination varies significantly.

    Since Encke gap is only 320~km wide, particles cannot have eccentricities larger than $1.2 \times 10^{-3}$ ($ae \approx 160 \km$) otherwise they would collide with the gap edges. If the eccentricity evolution is dominated by solar radiation pressure, from Eq.~\eqref{eq:forced_e_2} we conclude that in the Encke central ringlet only particles with radii larger than 2.4$\mum$ ($Q_{pr}\approx 0.3$) develop maximal eccentricities that are smaller than the critical eccentricity $1.2 \times 10^{-3}$, and thus, have the chance to survive the first period of the eccentricity evolution ($\sim$220\,days) described by equation~\eqref{eq:e}. For this reason one expects that particles smaller than 2.4$\mum$ are depleted in the Encke central ringlet. This is consistent with the interpretation of certain near infrared spectral features of the Encke gap ringlets in terms of a depleted fraction of few-micron-sized particles \citep{HedmanNicholsonEtAl2011}.

\subsubsection{Plasma drag}
    \label{sec:plasma}
    \paragraph{The plasma environment of the Encke gap}
    To calculate the forces exerted on a dust particle by ambient plasma, as well as the equilibrium surface potential of particles, one needs to know the plasma environment in the Encke gap.
    During Saturn Orbit Insertion (SOI) of the Cassini spacecraft in 2004, the plasma instruments onboard Cassini identified $\Op$ and $\Opp$ ions above Saturn's main rings between $1.8\,R_s$ to $2.05\,R_s$ \citep{YoungBerthelierEtAl2005,TokarJohnsonEtAl2005,WaiteCravensEtAl2005} and electrons \citep{CoatesMcAndrewsEtAl2005}.
Based on observation and modeling, the ion densities in the equatorial plane were estimated \citep[Fig. 4 of][]{JohnsonLuhmannEtAl2006}, along with the respective ion temperatures \citep{TokarJohnsonEtAl2005,JohnsonLuhmannEtAl2006}.

For the current modelling we assume that the ion densities in the Encke gap ($\sim 2.2 R_s$) are similar to those in the Cassini Division and the temperature can be described by $\textrm{E}_* + \textrm{E}_p$ \citep{JohnsonLuhmannEtAl2006}, where E$_*$ is the energy when ions are formed and E$_p$ is the ion pick-up energy. For the Encke gap this implies for the ion number densities $n_{\Op}=1\cm^{-3}$ and $n_{\Opp}=2\cm^{-3}$, and ion temperatures of $T_{\Op}=2.5$~eV and $T_{\Opp}=4.2$~eV. For electrons near Saturn's A ring, the observed temperatures $T_e$ range from 0.1 to 10~eV, and densities $n_e$ are 1--5$\cm^{-3}$ \citep{WahlundBostromEtAl2005_GRL_RPWS,CoatesMcAndrewsEtAl2005}. In the Encke gap we adopt an electron temperature of 2~eV and  3~cm$^{-3}$ for the electron density, assuming $n_e=n_{\Op} + n_{\Opp}$. 

\citet{TsengJohnsonEtAl2013} simulated the formation and transport of main ring neutrals and ions and suggest that the neutral and ion densities ($\O2$, $\Op$ and $\Opp$) near the main ring vary by almost 2 orders of magnitude during one Saturn year. In their simulation, the maximum value of $n_{\Opp} \approx 250 \cm^{-3}$ occurred in the Cassini Division when the solar incidence angle is large \citep{TsengIpEtAl2010_ica}. The ion densities are higher in the Cassini Division than in other regions of the main rings because there are fewer ring particles in the equatorial plane acting as sinks for the plasma particles (Tseng 2012, private communication). The densities from these simulations are about two orders of magnitude higher than those from the analytical model \citep{JohnsonLuhmannEtAl2006}.
For this study we only consider the case when ion densities are about a few $\cm^{-3}$.

In summary, in the following we will use fixed ion densities and temperatures $n_{\Op}=1\cm^{-3}$, $n_{\Opp}=2\cm^{-3}$, $T_{\Op}=2.5$~eV and $T_{\Opp}=4.2$~eV, and electron density $n_e = 2 \cm^{-3}$ and electron temperature $T_e = 3$~eV. These values are used to calculate the surface potential of grain and simulation of particle dynamics (the modified horseshoe orbit). Larger ion density ($n_{\Opp} = 30 \cm^{-3}$) is also used in the demonstration of the modified horseshoe orbit.

    \paragraph{Particle charging}
    The evolution of electric charge $q$ of a particle is quantified by \citep{Horanyi1996}
    \begin{equation}
	\label{eq:charging}
      \frac{dq}{dt} = \sum_k J_k
    \end{equation}
    where $J_k$ are the charging currents. These are the photo-current due to Solar UV, absorption of ionospheric electrons and ions, and secondary electron emission. Detail on the charging kinetics can be found in \citet{Horanyi1996}. In the Encke gap, particles are dominantly charged by sweeping up electrons and ions, the crucial parameters are temperatures and densities of electrons. In equilibrium all currents on a grain balance, giving $\dot{q} = 0$, and one obtains the equilibrium grain charge from application of formula (\ref{eq:charging}).  The equilibrium charge is related to the free floating potential $\phi$ of a particle in a given charging environment by \citep{HoranyiBurnsEtAl1992}
    \begin{equation}
        \phi \approx 300 \, q / s
    \end{equation}
    i.e.,\ evaluating the Coulomb potential on the grain surface at radius $s$ (CGS units).
    The surface potential is independent of the grain radius, in other words, larger grains and smaller grains eventually reach the same surface potential.
    Applying the afore mentioned electron and ion properties for the Encke gap, one obtains a grain surface potential of about -4~V.
    Note that the surface potential could range from 0~V to -15~V by varying the temperatures and densities in the observed range ($\phi \sim 0$~V when $T_e<0.1$~eV, -15~V is when $T_e \approx 10$~eV), where the result is quite sensitive to electron temperatures.

    \paragraph{Plasma drag force}
    Plasma interacts with particles via Coulomb interaction between charged particles of bulk of plasma ($\mathbf{F}_{C}$) and direct collisions ($\mathbf{F}_{D}$).
    For Maxwellian plasma, the Coulomb drag can be written as (\citealt{NorthropBirmingham1990}, in CGS units)
    \begin{equation}
      \begin{split}
      \label{eq:Fc}
      F_C = \frac{\sqrt{\pi} q^2 \tilde{e}^2 n_i}{m_i \tilde{v}_0^2}
            & \int^{\infty}_{-\infty} \frac{y}{\abs{y^3}}(2M_i y - 1) \exp[-(y-M_i)^2] \\
            & \ln{\frac{1+(m_i u_i^2 \lambda_D y^2 / (q\tilde{e}))^2}{1+(m_i u_i^2 s y^2 / (q\tilde{e}))^2}}dy
      \end{split}
    \end{equation}
    whereas for the direct collisions \citep{BanaszkiewiczFahrEtAl1994}
    \begin{equation}
      \begin{split}
        \label{eq:Fd}
        F_D = \pi n_i m_i u_i^2 s^2
            & \bigg[  (M_i + \frac{1}{2 M_i}) \frac{\exp(-M_i^2)}{\sqrt{\pi}} + \\
            & (M_i^2 + 1 - \frac{1}{4 M_i^2}) \erf(M_i) \bigg].
      \end{split}
    \end{equation}
    The particles are defined by their radii $s$ and charge $q$.
    The ions are characterized by their number density $n_i$, thermal velocity $u_i$, mass $m_i$, temperature $T_i$, and charge $\tilde{e}$, respectively.
    The Mach number is defined by $M_i = \tilde{v}_0/u_i$, where $\tilde{v}_0$ is the velocity of a particle relative to bulk of plasma.
    The Debye length is given by $\lambda_D = (kT_i / 4 \pi n_i \tilde{e}^2)^{1/2}$.
    The integration in Eq.~\eqref{eq:Fc} is with respect to ion speed $y$ (normalized by ion thermal speed $u_i$) while using a Maxwellian velocity distribution.

\begin{figure}
        \begin{center}
            \resizebox{0.62\textwidth}{!}{\includegraphics{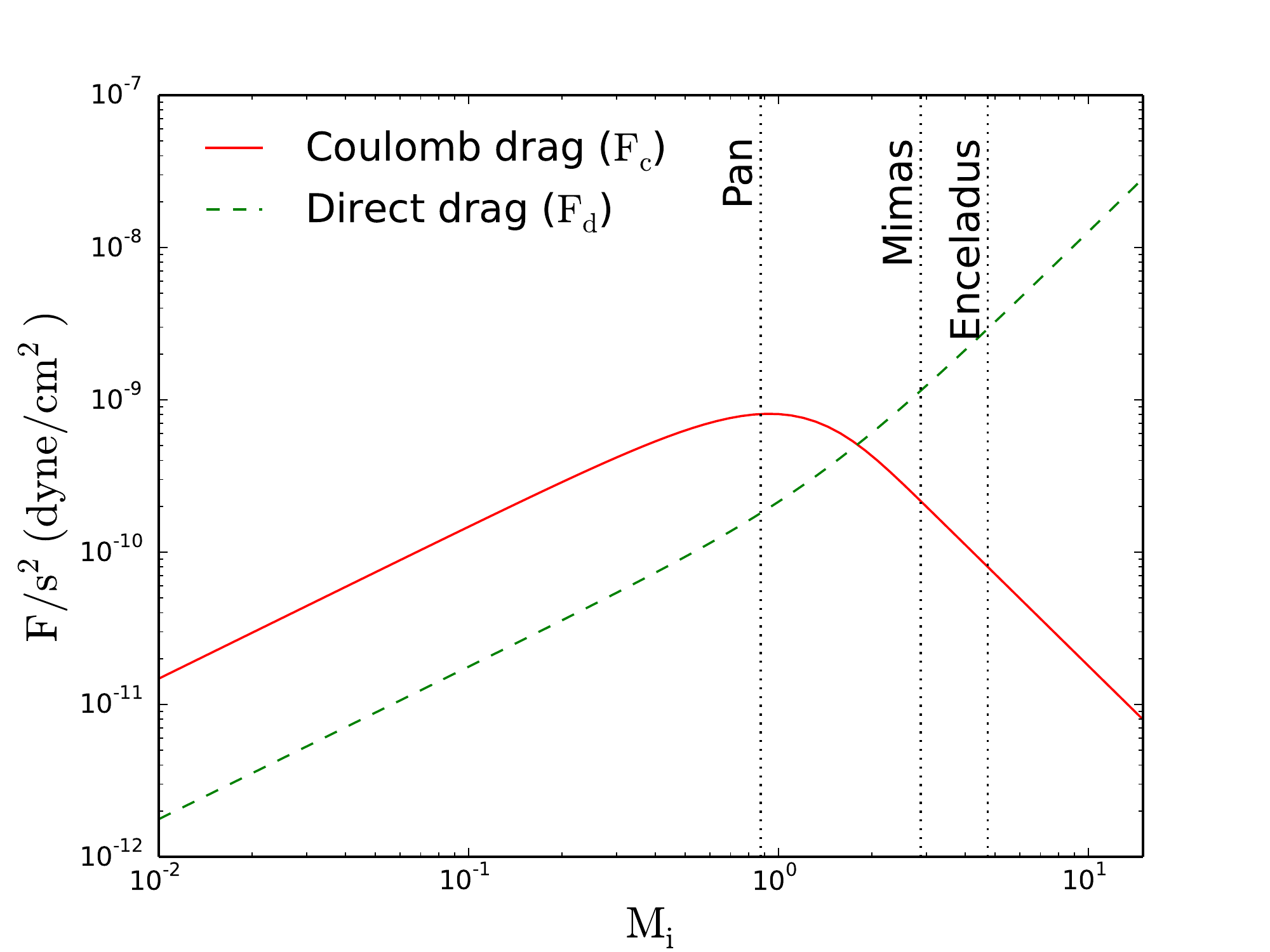}}
            \caption{Pressure act on charged particles with different Mach numbers $M_i$.
            The values are calculated by using the plasma environment described in the text for the Encke gap ($n_{\Op}=1\cm^{-3}$, $n_{\Opp}=2\cm^{-3}$, $T_{\Op}=2.5$~eV, and $T_{\Opp}=4.2$~eV), and the grain radius is 10$\mum$ with constant surface potential of -4~V.
            The three vertical dashed lines marked the Mach numbers for grain near Pan, Mimas, and Enceladus.
            For this figure, only the \emph{relative} strength of two plasma drag at different $M_i$ is important, not the absolute values.
            Note that in comparison to the similar figure in \citet[fig. 6]{GrunMorfillEtAl1984}, we adopt the $F_D$ from \citet{BanaszkiewiczFahrEtAl1994} which consider collisions of ions from different angles.
            }
            \label{fig:Fcd}
        \end{center}
    \end{figure}

    In the Encke gap, using the plasma environment described above and the surface potential of particle -4~V, Eqs.~\eqref{eq:Fc} and \eqref{eq:Fd} show that Coulomb drag $F_{C}$ is several times stronger than the plasma direct collision term $F_{D}$ (see Fig.~\ref{fig:Fcd}).
    This differs from the situation in Saturn E ring (the inner edge is at 3 $R_s$, about $M_i \approx 2.8$ in Fig.~\ref{fig:Fcd}), where $F_{D}$ dominates and $F_{C}$ is ignorable \citep{Dikarev1999,JuhaszHoranyiEtAl2007_e_ring}.
    In the following, we consider both drags and refer plasma drag as the combination $\mathbf{F}_{C}+\mathbf{F}_{D}$.

    At inner magnetosphere of Saturn, plasma corotates with Saturn's magnetic field. The synchronous orbit of Saturn is at $\sim$1.86~$R_s$, it means particles in nearly circular orbits at Encke gap ($\sim$2.2~$R_s$) have orbital speeds smaller than that of plasma.
    Hence (negatively charged) particles are accelerated along their orbital motion by both direct collision and Coulomb interaction.
    The change of orbital elements can be estimated by perturbation equations (Eq.~\eqref{eq:perturba}-\eqref{eq:perturbW}). Assuming the eccentricity of particle is zero, the plasma drag only acts on the transverse direction of particle orbit, i.e, $\mathbf{F}_R = \mathbf{F}_N = 0$ and $\mathbf{F}_T = \mathbf{F}_C + \mathbf{F}_D$.
    Applying the corresponding values, we obtain a growth rate of semi-major axes of dust grains in the Encke gap
    \begin{equation}
        \label{eq:dadt}
        \frac{da}{dt} \approx 1 \left[\frac{10\mum}{s}\right] \quad \km/\yr \,,
    \end{equation}
    where the perturbations on other orbital elements are negligibly small.
    Such outward migration rate implies that a 10$\mum$ particle in the Encke central ringlet needs only $\sim$19 years to leave the ringlet and $\sim$160 years to reach the outer edge of the Encke gap. Fortunately, central ringlet particles are also confined in the horseshoe region of Pan therefore plasma drag is not destroying the ringlet in the timescale of decades. The combination of the outward migration and horse-shoe orbit, or the modified horseshoe orbit, will be discussed in Section~\ref{sec:plasma_hs}.

    As we mentioned earlier that ion densities in main ring ionosphere could be 1--2 orders of magnitude larger than the value we adopted here. Since both plasma drags are proportional to ion densities (Eq.~\eqref{eq:Fc} and \eqref{eq:Fd}), this indicates that these forces could be 10--100 times larger than expected here.

\subsubsection{Summary of perturbing forces}
    \label{sec:sum_dyn}
\begin{figure}
        \begin{center}
            \resizebox{0.95\linewidth}{!}{\includegraphics{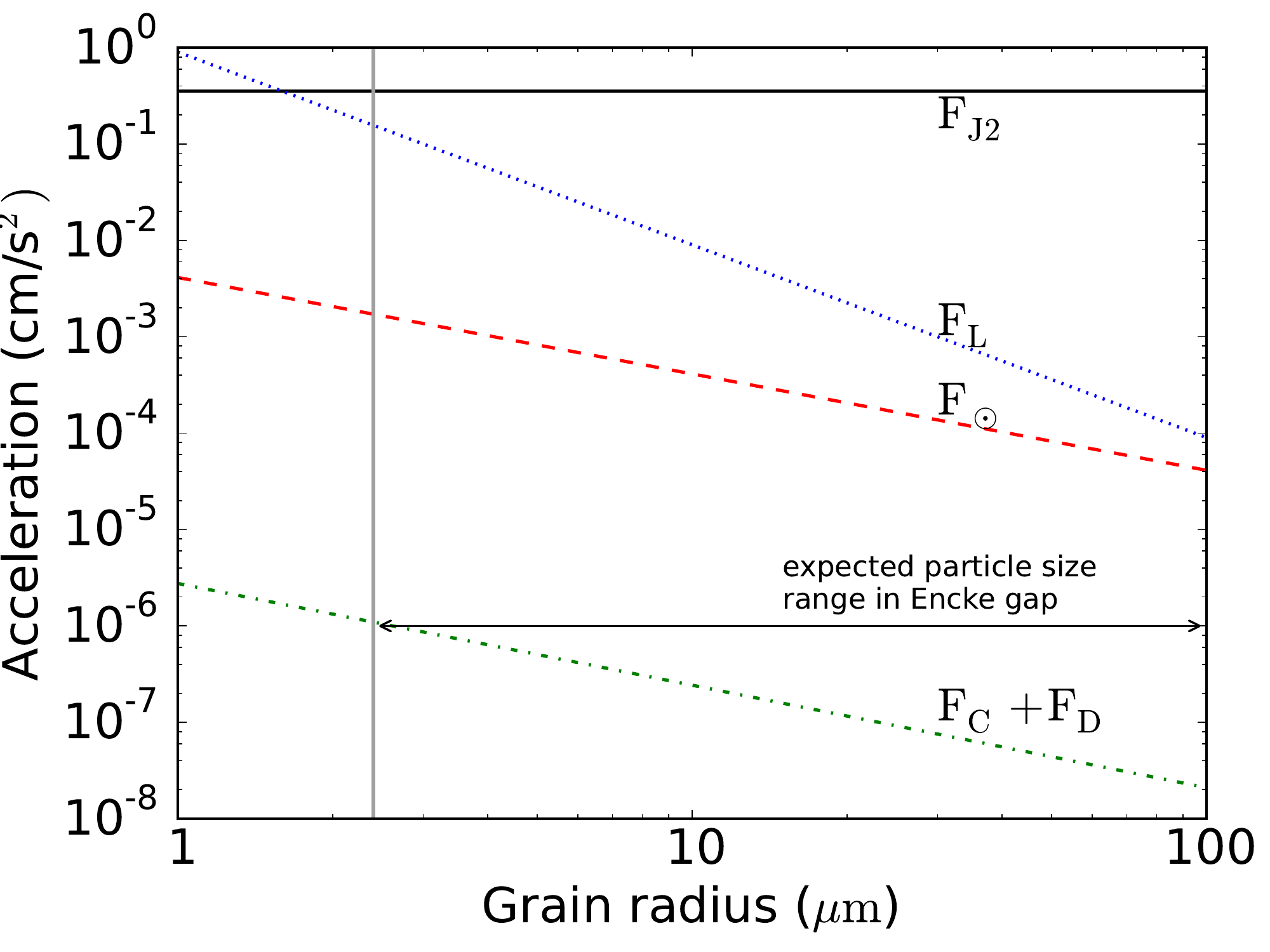}} 
            \caption{Comparison of perturbation forces for particle size range from 1 to 100$\mum$
            in Encke gap. $\mathbf{F}_{J2}$ is the perturbation force of Saturn's oblateness.
            Assuming the particles are in near circular orbit with
            surface potential of -4 V, and the ions are $\Op/\Opp$ with temperature 2.5/4.2 eV
            and number density 1/2$\cm^{-3}$.
            In Encke gap, we conclude that the particle size is in the range of 2.4--100$\mum$.
            }
            \label{fig:comparison_perturbation}
        \end{center}
    \end{figure}

    The strength of perturbing forces for Encke ringlet particles of radius 1--100$\mum$ with initially nearly circular orbits, relevant for the description of the Encke gap ringlet particles (2.4--100$\mum$), is shown in Fig.~\ref{fig:comparison_perturbation}.
    Below we summarize the particle orbits perturbed by these forces and it's implications.
    \begin{itemize}
        \item The precession rate of pericenter and ascending nodes are caused by the oblateness of Saturn and the Lorentz force \citep{HoranyiBurnsEtAl1992}. As shown in Fig.~\ref{fig:comparison_perturbation}, the latter becomes ignorable for particles larger than a few micrometer. Since observations and the forced eccentricities both suggested that the Encke ringlets are lack of few-micron-sized particles, Lorentz force can be ignored.
        \item The eccentricities and inclinations are perturbed by solar radiation pressure with shadow of Saturn considered \citep{HedmanBurtEtAl2010}. It restricts the minimum particle size to 2.4$\mum$ in Encke central ringlet. Observations also suggest that Encke ringlet lack of few-micron-particles \citep{HedmanNicholsonEtAl2011}.
        \item The semi-major axes are increased at a rate of about 10--0.1 km/yr (for particles range from 1--100$\mum$) due to plasma drag. The increase of $a$ can accumulate over time, therefore although the strength of the plasma drag is relatively weak, it must be considered.
    \end{itemize}

\subsection{Sinks}
    \label{sec:sinks}
    The last ingredient of the kinetic model concerns the sinks of ringlet particles, which are
    collision (with moonlets, Pan, and the A ring) and erosion.
    In the following we demonstrate that the former dominates in the presence of an embedded moonlet belt of optical depth $\tau_m = 10^{-6}$.
    Colliding with gap edges also removes particles from Encke central ringlet. But as demonstrated earlier (Section~\ref{sec:F_rad}), this mainly removes particles with radii smaller than 2.4$\mum$ in very short time scales (a few particle orbits). Below we only consider particles larger than that size.

\subsubsection{Collision with moonlets}
    \label{sec:sinks_moons}
    The collisional lifetime $T_{col}$ of a particle among $k$ moonlets can be quantified by \citep{MakuchKrivovEtAl2005}
    \begin{equation}
        \label{eq:t_col_life}
        \frac{1}{T_{col}} = \sum_{j=1}^{k} \frac{1}{T_{j}}
    \end{equation}
    where $T_j$ is collisional lifetime of the particle with $j$th moonlet and can be estimated by using orbital elements $a$, $e$, and $i$ of the particle and $j$th moonlet with the approach developed by \citet{Opik1976}
    \begin{equation}
        \label{eq:T_moon}
        T_{moon} \approx \pi \sqrt{\sin^2{i} + \sin^2{i_{moon}}}
            \left( \frac{a_{moon}}{R_{moon}} \right)^2
            \left( \frac{u}{u_r} \right) P
    \end{equation}
    where $R_{moon}$, $a_{moon}$, and $i_{moon}$ are moonlet radius, semi-major axis, and inclination, $P$ is the orbital period of the particle, $u$ is the average particle velocity relative to the moonlet and $u_r$ is the radial component of $u$. The relative velocities are expressed as
    \begin{equation}
        u = \sqrt{3 - \frac{1}{A} - 2 \sqrt{A(1-e^2)}}
    \end{equation}
    \begin{equation}
        \label{eq:T_moon_ur}
        u_r = \sqrt{2 - \frac{1}{A} - A(1-e^2)}
    \end{equation}
    with $A \equiv a/a_{moon}$.
    In the case of a parent moonlet of radius 250 meter embedded in the ringlet with small ($10^{-5}$) eccentricities and inclinations, the collisional lifetime of a 5$\mum$ ringlet particle (with median value of $e = 1.7 \times 10^{-3}$ and $i = 2.3 \times 10^{-5}$ induced by solar radiation pressure, see Eqs.~\eqref{eq:forced_e_2}-\eqref{eq:forced_i_2}) is about 36,000 years, or about 360 years if there is a moonlet belt containing 100 similar objects. Larger particles have shorter lifetimes, due to smaller $e$ and $i$: for 10$\mum$ the lifetime is about 217 years, and for 20$\mum$ it is 160 years.

    This analytical approach is sensitive to the particle and moonlet orbits, and therefore it can just be used for first order estimates.
    In Section~\ref{sec:results} we present the particle lifetimes obtained by simulations which consider the essential perturbing forces. It gives $\angles{T_{moon}}$ in the order of decades dependent on particle size. The big difference in the model and simulation result is likely because the analytical approach does not consider the time evolution of particle orbits, especially eccentricities and inclinations.

\subsubsection{Plasma sputtering}
    Apart from collision, lifetimes of icy particles near Saturn are also restricted by erosive processes, mainly by sputtering due to magnetospheric plasma.
    Other erosive processes, such as sublimation by solar UV radiation or micrometeoroid shattering, require timescales of $10^5$ years or more \citep{Burns2001_inbook}. 
    The plasma sputtering rate is roughly in the form $\phi_{sput} \propto \tilde{v}_0 n_i Y_i(E_i)$, where $\tilde{v}_0$ is the speed of the particle relative to bulk of plasma, $n_i$ is ion density, and $Y_i(E_i)$ is the yield (averaged over impact angles) dominated by ion energy $E_i$ \citep{JohnsonFamaEtAl2008_sputtering}.
    \citet{JohnsonFamaEtAl2008_sputtering} gives the sputtering rate $\phi_{sput} \sim 10^9 \H2O\cm^{-2}\s^{-1}$ for water ice particle exposed to plasma environment at 4~$R_s$, which means the erosive lifetime is 100 years for grain with 1$\mum$ radius.
    In the Encke gap ($\sim 2.2\,R_s$), $n_i$ is of the same order (a few $\cm^{-3}$, see Section~\ref{sec:plasma}) and both $\tilde{v}_0$ and $Y_i(E_i)$ are smaller. 
    Hence we adopt a smaller erosion rate, $ \phi_{sput} \sim 10^8 \, \H2O\cm^{-2}\s^{-1}$, which gives a rough estimate to the plasma sputtering timescale 
    \begin{equation}
        \label{eq:sputter}
        T_{sput} \approx 10^4 \left( \frac{s}{10 \mum} \right) \, \mathrm{years} \,.
    \end{equation}
    Therefore, the lifetime of Encke central ringlet particles is about several decades to a century -- dominated by collisional processes.

\subsection{Summary of kinetics}
    \label{sec:sum_kin}
    We have introduced the kinetic balance of ringlet, concerning the birth, evolution, and death processes (sources, dynamics, and sinks) of ringlet particle.
    Our major assumption is that the sources and sinks are related to a putative moonlet belt with total optical depth $\tau_m \approx 10^{-6}$, which correspond to the surface area of 100 moonlets with radius 250 meter.
    With this assumption, we then estimate the dust production rate caused by impact-ejecta model \citep{KrivovSremvEtAl2003,SpahnAlbersEtAl2006} summarized in Table~\ref{tbl:n_plus}.
    We also introduce other mechanisms to produce dust, such as secondary sources \citep{DikarevKrivovEtAl2006} and mutual collisions between moonlets \citep{BodrovaSchmidtEtAl2012,CanupEsposito1995}. The former enhance the impact-ejecta process by factor of about 10, while the latter probably provide same amount of dust as impact-ejecta process but only on the base of some poorly constrained parameters.

    Once particles are released, their orbital evolution are controlled by perturbing forces. For example, the solar radiation pressure induce particle eccentricities which is inverse proportional to particle size. It indicates the existence of a minimum size for particles, since smaller ones quickly collide with gap edges in very short timescale (few orbital periods). The heliotropic behavior is also induced by solar radiation pressure \citep{HedmanBurtEtAl2010}. A further important perturbation is the plasma drag, which causes the ringlet particles to migrate outward at a rate of about 1~km/year (for 10$\mum$ radius particles).

    To reach a kinetic balance, particle sinks are as important as the sources.
    With our main assumption of a putative moonlet belt, the particle lifetimes are dominated by their collisions with the these moonlets.
    The collisional lifetime of particles is estimated by the \citet{Opik1976} approach, giving lifetimes about 360 years, which is larger than the result of numerical simulations ($\sim$30 years) discussed below.
    We also show that the erosion lifetime of icy ringlet particles is about $10^4 (s/10\mum)$ years due to plasma sputtering.

\section{Simulation results and discussion}
    \label{sec:results}
    Simulation results are divided into two major parts: (1)~an averaged lifetime $\angles{T_{life}}$ is found; (2)~the examples of modified horseshoe orbits caused by gravitational forces and plasma drag.
    Numerical integration of Eq.~\eqref{eq:motion} are done by using the RADAU integrator \citep{Everhart1985} implemented by \citet{Chambers1999}.

    Since most perturbing forces are size-dependent, we use different particle radii of 5, 10, and 20$\mum$ in the simulations. Note that the averaged particle size $\angles{s}$ is about $5\mum$ (see Section~\ref{sec:source}).

\subsection{Impact-ejecta: Pan as source}
    \label{sec:res_Pan}
    We already exclude Pan as source of Encke central ringlet (in Section~\ref{sec:source}) due to the fact that ejecta from Pan are gravitationally scattered by Pan to another orbits. Although Pan produces more particles than the putative embedded moonlets, it cannot be an efficient source of the Encke ringlet itself. Below we use numerical simulations to demonstrate this.

    As listed in Table~\ref{tbl:n_plus}, the optical depth increase rate by Pan $\dot{\tau}^+ = 2.9 \eta \times 10^{-7}$, where $\eta$ is the fraction of ejecta from Pan that can contribute to the Encke central ringlet.
    To quantify $\eta$, we numerically simulate 8,000 test particles ejected from Hill sphere of Pan with practically zero initial velocity. Only 10 particles have semi-major axes in the range of Encke central ringlet. In other words, $\eta = 0.125\%$ and therefore $\dot{\tau}^+ = 3.6 \times 10^{-10}$. Due to plasma sputtering, even if there is no other embedded moonlets act as sinks, lifetimes of 10-micron-particles cannot exceed 10,000 years, this means a ringlet merely sustained by ejecta from Pan can only have optical depth less than $3.6 \times 10^{-6}$. This is much smaller than the observed maximum optical depth 0.1.

\subsection{Impact-ejecta: Embedded moonlets as source}
    \label{sec:res_ml}
    Now back to our main assumption that there are 100 moonlets with radius 250~meter in the Encke central ringlet act as dust sources and sinks.
    In this model, the life cycle of a particle is: it is ejected from a moonlet through impact-ejecta process; it stays in the horseshoe orbit of Pan (like the source moonlets) while the perturbing forces are additionally at work; finally, the particle is removed from ringlet due to collision with an embedded moonlet. Other particle sinks such as plasma sputtering or collision with gap edge are both not necessary to consider here: the former require a much larger timescale ($\sim10^4\yr$), while the latter should only happen on smaller ($s<2.4\mum$) particles.
    In Section~\ref{sec:source} we already derive the optical depth increase rate by $\dot{\tau}^+ = 3.4 \times 10^{-8} \yr^{-1}$, in the following we numerically estimate the averaged particle lifetime $\angles{T_{life}}$ in order to obtain a steady state density of the ringlet.

    In Section~\ref{sec:sinks} we estimate the averaged particle lifetime $\angles{T_{life}}$ by the analytical model of \citet{Opik1976}, which is very sensitive to the orbital parameters $e$ and $i$ of the particle and therefore may only be used as first order estimates. There we obtain $\angles{T_{life}}$ in the order of a few hundred years.
    Here we evaluate $\angles{T_{life}}$ by numerical simulations.
    In these simulation, moonlets are assumed to be massless and evenly distributed over the ringlet and revolve in horseshoe orbits with zero eccentricities and inclinations in the central ringlet. The neglect of the moonlets' gravity are motivated by the fact that they should have filled their Hill spheres. Thus, they don't act as efficient scatters but rather as sinks of dust. 
    Particle radii of 5$\mum$, 10$\mum$, and 20$\mum$ have been chosen for simulations.
    For each of the sizes, 6,400 particles are launched simultaneously from the Hill spheres of moonlets with practically zero velocity (the initial ejecta speed are small, as discussed in Section~\ref{sec:source}).

    Once particles are ejected, the gravity forces (including Saturn oblateness) and solar radiation pressure are taken into account.
    Lorentz force and plasma drag are ignored in these simulations.
    Lorentz force is too weak for particles larger than a few $\mum$, as discussed in Section~\ref{sec:grav_lore}.
    The plasma drag increase the particle semi-major axes slowly and is ignored here -- but will be addressed in Section~\ref{sec:plasma_hs}. The reason is that collisional lifetime is more sensitive to particle eccentricities and inclinations than to small semi-major axis variations (Eq.~\eqref{eq:T_moon}--\eqref{eq:T_moon_ur}).
    Particles are discarded once they collide with the embedded moonlets. Other sinks have been found to be negligible.

\begin{figure}
        \centering
        \begin{subfigure}
            \centering
            \includegraphics[width=0.49\textwidth]{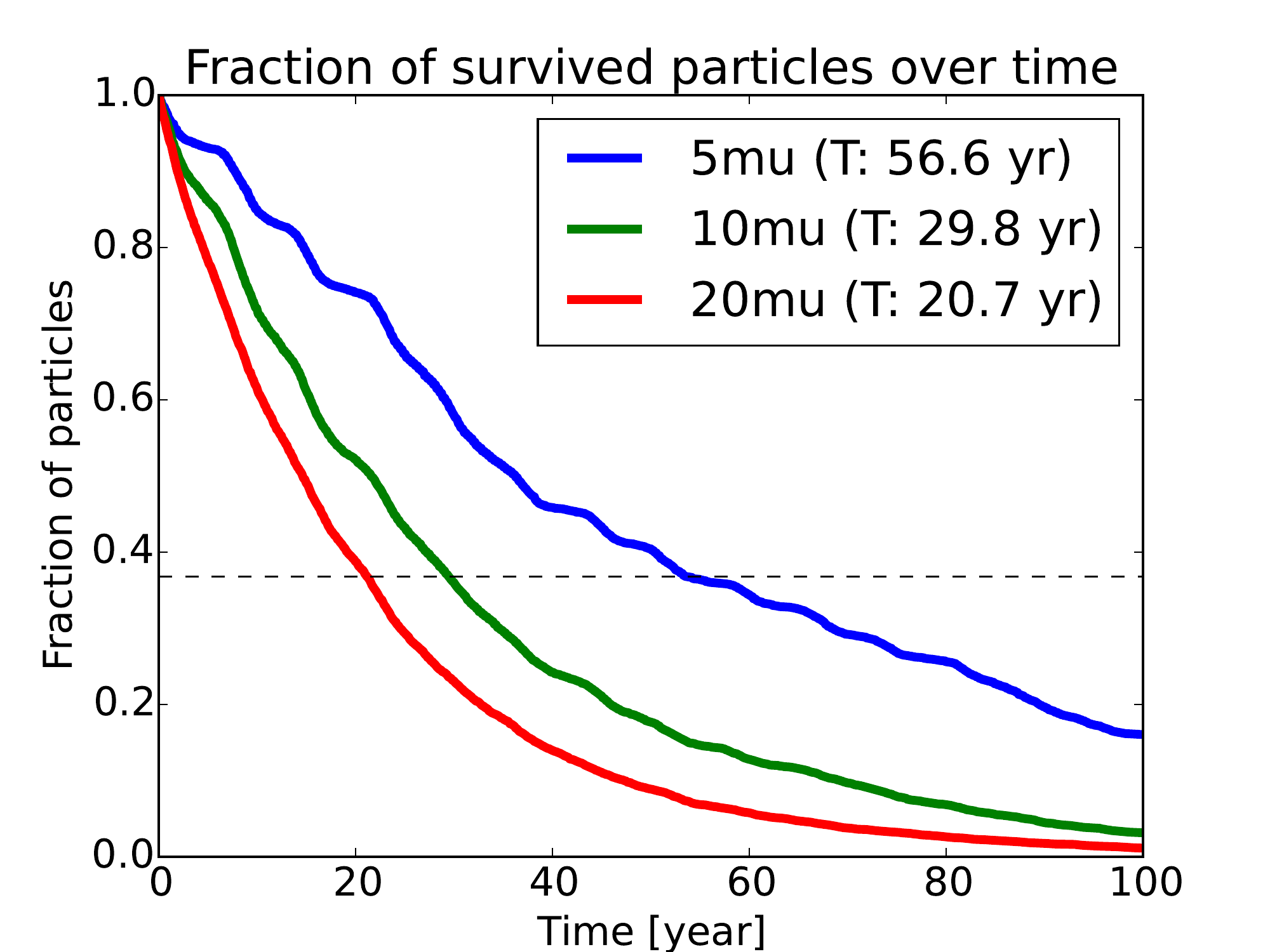}
        \end{subfigure}
        \begin{subfigure}
            \centering
            \includegraphics[width=0.49\textwidth]{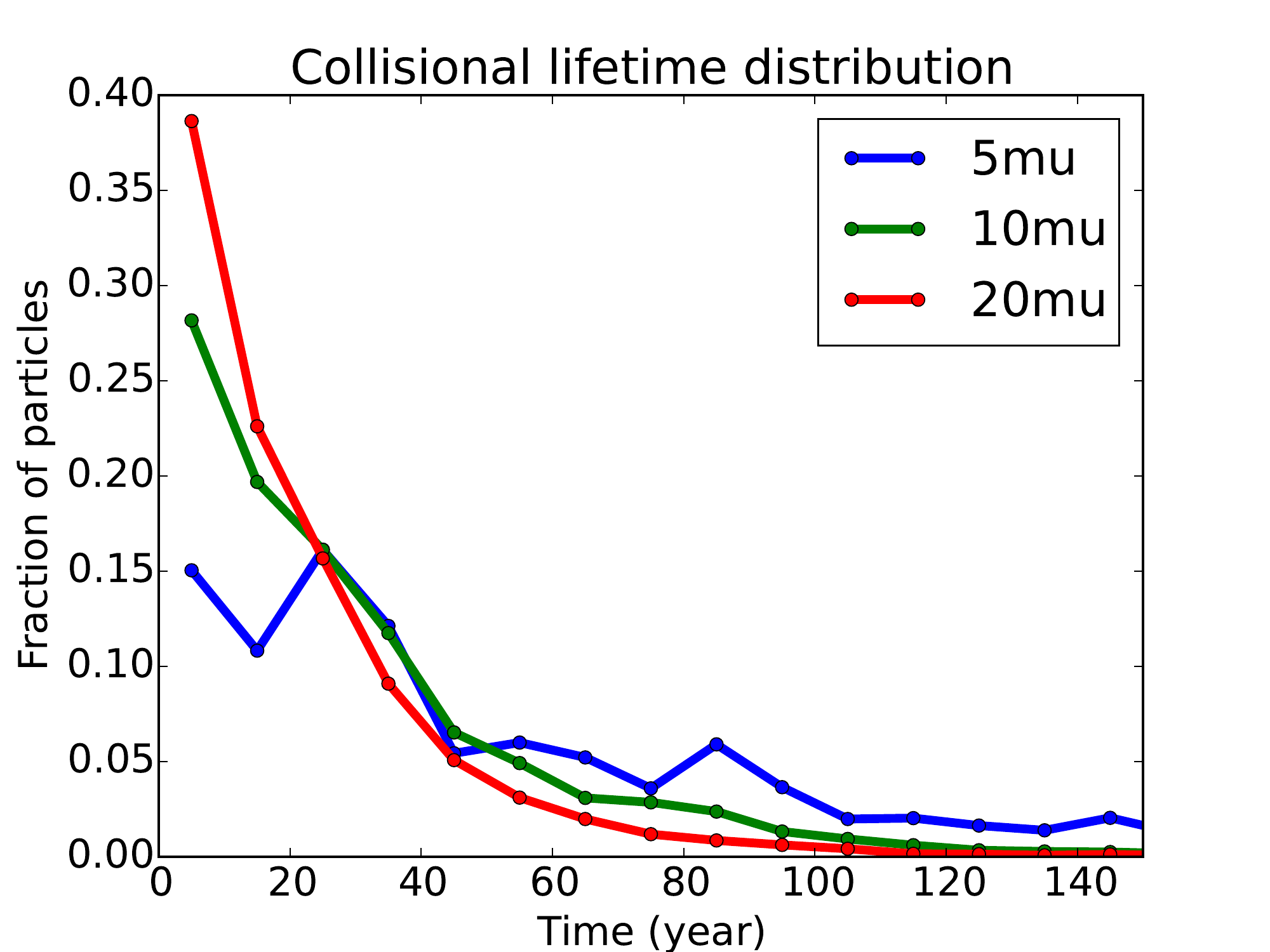}
        \end{subfigure}
        \caption{Simulation results on the timescales for 5/10/20$\mum$-radius particles to collide with embedded moonlets. There are 6400 particles for each size. The horizontal dashed line on the left panel marks $\exp(-1)$, and the intersections of this dashed line and curves give the averaged particle lifetime $t = \angles{T_{life}}$ for each curve.
            \textit{Left}: Fraction of surviving particles over time.
            \textit{Right}: The particle lifetime distributions.
        }
        \label{fig:lifetime}
    \end{figure}

    The fraction of surviving particles and particle lifetime distributions of each selected particle sizes in the simulation is shown in Fig.~\ref{fig:lifetime}.
    The fraction of survived particles can be roughly described by $\exp (-t/\angles{T_{life}})$.
    Therefore, the $\angles{T_{life}}$ for 5/10/20$\mum$-radii particles are around 56.6/29.8/20.7 years.

    Finally, with the optical depth increase rate $\dot{\tau}^+ = 3.4 \times 10^{-8} \yr^{-1}$ (Table~\ref{tbl:n_plus}) and $\angles{T_{life}}=56.6 \yr$ (the lifetime of particles with averaged radius $\angles{s} = 5\mum$), we obtain $\tau \approx 1.92 \times 10^{-6}$.
    This result is about 4 orders of magnitude smaller than the observation.

\subsubsection{Enhancement of the impact-ejecta model}
    \label{sec:extension}
    The resulting optical depth of impact-ejecta process is several orders of magnitude lower than observation, indicating that something crucial is missing in the model.
    The `missing' methods must be increase either the particle production rate or the particle lifetime.
    As discussed in Section~\ref{sec:second}, the impact-ejecta could form a dust belt which itself serves as secondary sources of dust through mutual collision and micrometeorite impact \citep{DikarevKrivovEtAl2006}. The enhancement factor ($D_{int} =  \dot{N}_{int}/\dot{N^+}$) is about 9.7.

    Another possibility is that we have underestimated the particle lifetime due to missing physical effects. For example, the simulation result in Fig.~\ref{fig:lifetime} shows that smaller particles have longer lifetimes -- probably because their larger $e$ and $i$ induced by solar radiation pressure cause their orbit more deviated from the embedded moonlets.
    So, if the ejecta size distribution is steeper and therefore the average radius $\angles{s}$ is smaller, the lifetime may increase by a few decades.

    Another straightforward way to increase the dust production rate is to increase the total cross section of the moonlets $S$ (Eq.~\eqref{eq:mpr}).
    However, increase $S$ also decrease the particle lifetime in the same rate (Eq.~\eqref{eq:T_moon}).
    The ringlet will reach the same steady state (same optical depth) whether $S$ is larger or smaller than we expected. The exception is that when $S$ is too large or too small that the particle sources and sinks are simply not related to impact-ejecta process and collision with embedded moonlets.

    In conclusion, for impact-ejecta process, secondary sources and the extended lifetime due to plasma drag can increase optical depth by about one order of magnitude, i.e., $\tau \approx 2 \times 10^{-5}$. Still 3 orders of magnitude to be bridged in order to explain the observed Encke central ringlet.

\subsection{Modified horseshoe orbits}
    \label{sec:plasma_hs}
    Up to now, the plasma drag is ignored in the simulations which determine the particle lifetime for the sake of simplicity. Here we include plasma drag, together with gravity force of Saturn and Pan. The Encke central ringlet particles are then confined in so-called `modified' horseshoe orbits, resulting in a concentration of particles in a small range of longitudes in leading orbital direction of Pan.
    The idea that the combination of drag force and gravity of Pan may be related to the azimuthal brightness variation in the central ringlet has been proposed by \citet{Hedman2013}. To verify this idea, we simulate the particle orbital evolution with consideration of the Coulomb drag $\mathbf{F}_C$, which is several times larger than direct collision $\mathbf{F}_D$ in the vicinity of Encke gap. The other perturbing forces are also considered.

    \begin{figure}
        \begin{center}
            \resizebox{0.82\linewidth}{!}{\includegraphics[angle=0]{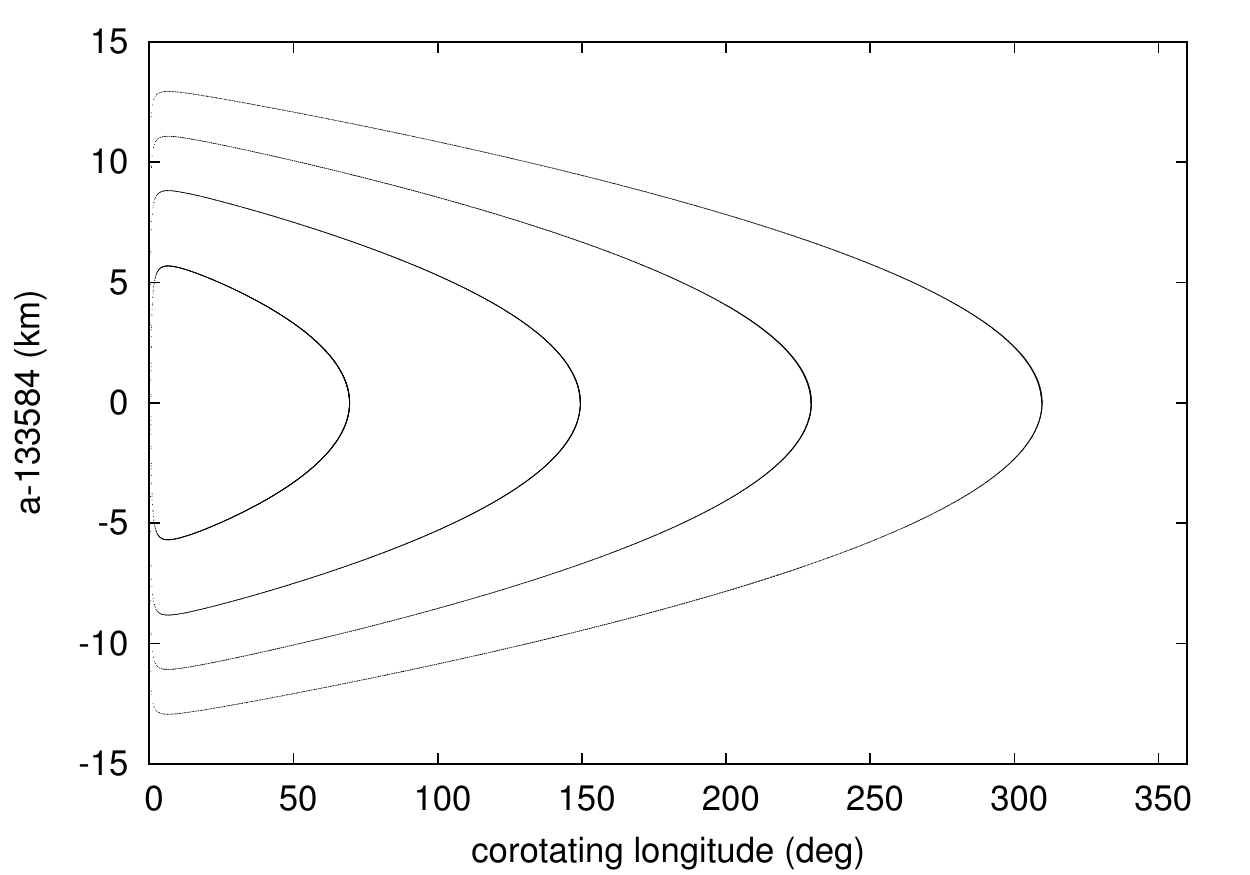}}
            \caption{Example of modified horseshoe orbit (simulation result). Pan is fixed at origin and the trajectories are 4 particles launched from different corotational longitudes relative to Pan and with same $a=a_{Pan}$. All trajectories are revolve counter-clockwise.
            }
            \label{fig:plasma_hs_3}
        \end{center}
    \end{figure}

    As a demonstration, simulated trajectories of 4 selected particles confined in the modified horseshoe orbits are shown in Fig.~\ref{fig:plasma_hs_3}. The trajectories are drawn in corotational frame with Pan.
    In the simulation we use the ion properties discussed earlier ($n_{\Op} = 1\cm^3$, $n_{\Opp} = 2\cm^3$, $T_{\Op}=2.5$~eV, and $T_{\Opp}=4.2$~eV) and use particles with radii $s= 20\mum$ and surface potential $\varphi=-4$~V.
    Particles are launched from semi-major axes of Pan ($a_{Pan}$) with different initial corotational longitudes ($70^\circ$, $150^\circ$, $230^\circ$, $310^\circ$). These particles are pushed outward to $a>a_{Pan}$ because of plasma drag. At $a > a_{Pan}$, the mean motion of particles become smaller than Pan and therefore they are moving towards smaller corotational longitudes (toward left of Fig.~\ref{fig:plasma_hs_3}); when they encounter with Pan, they behave like particles in horseshoe orbit that jump from $a_{Pan}+da$ to $a_{Pan}-da$, where $da$ is the accumulated outward migration. If the plasma drag does not vary significantly with time and location, these particles will eventually return to their launching positions and continue their cycles. It means that particles cannot reach large longitudes as ordinary horseshoe orbiting particles. Such an effect, first proposed by \citet{Hedman2013}, should corresponds to the observed azimuthal brightness variations in Encke central ringlet \citep{Hedman2013}.

\begin{figure*}
        \centering
        \begin{subfigure}
            \centering
            \includegraphics[width=0.48\textwidth]{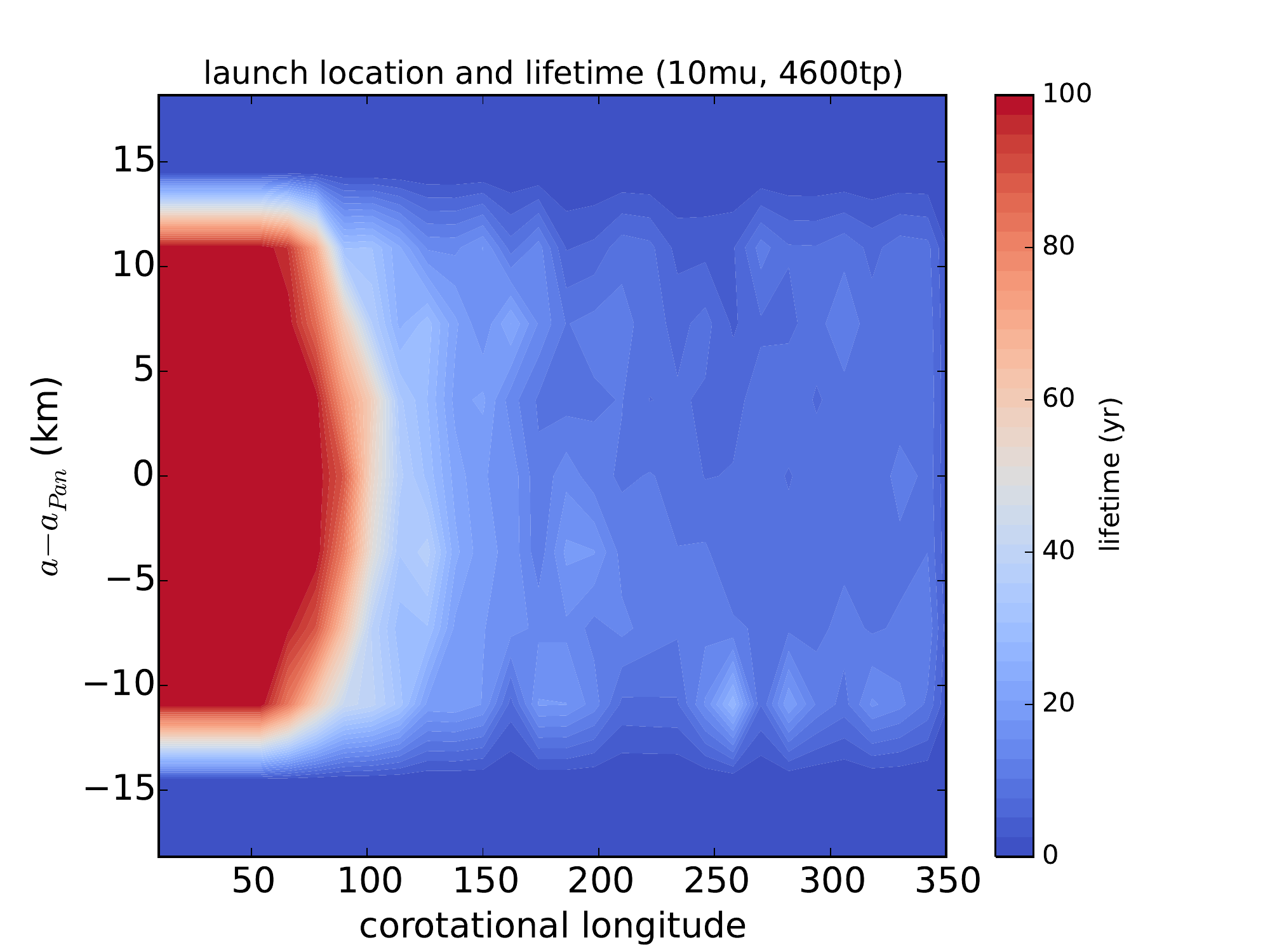}
        \end{subfigure}
        \begin{subfigure}
            \centering
            \includegraphics[width=0.48\textwidth]{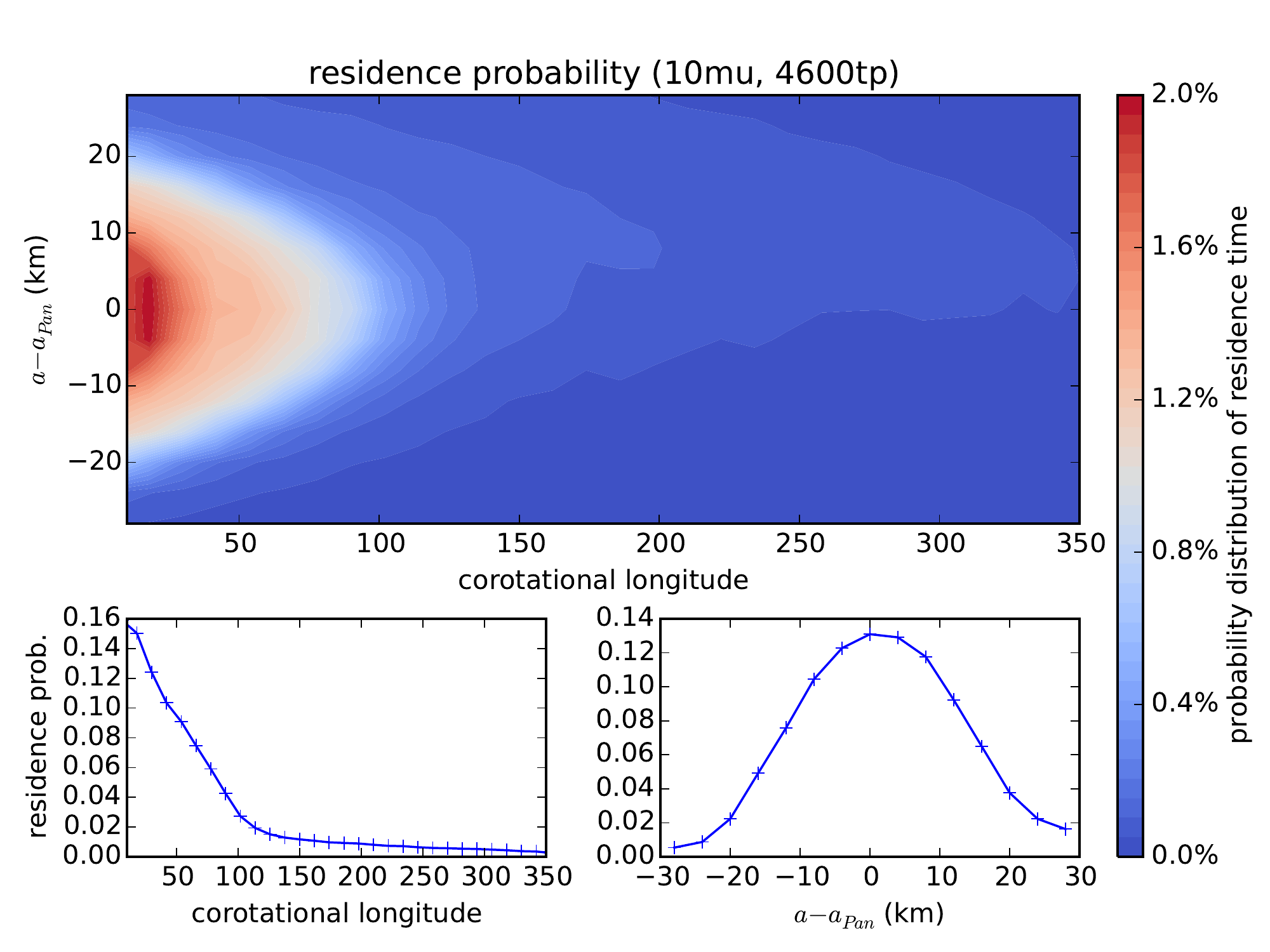}
        \end{subfigure}
        \begin{subfigure}
            \centering
            \includegraphics[width=0.48\textwidth]{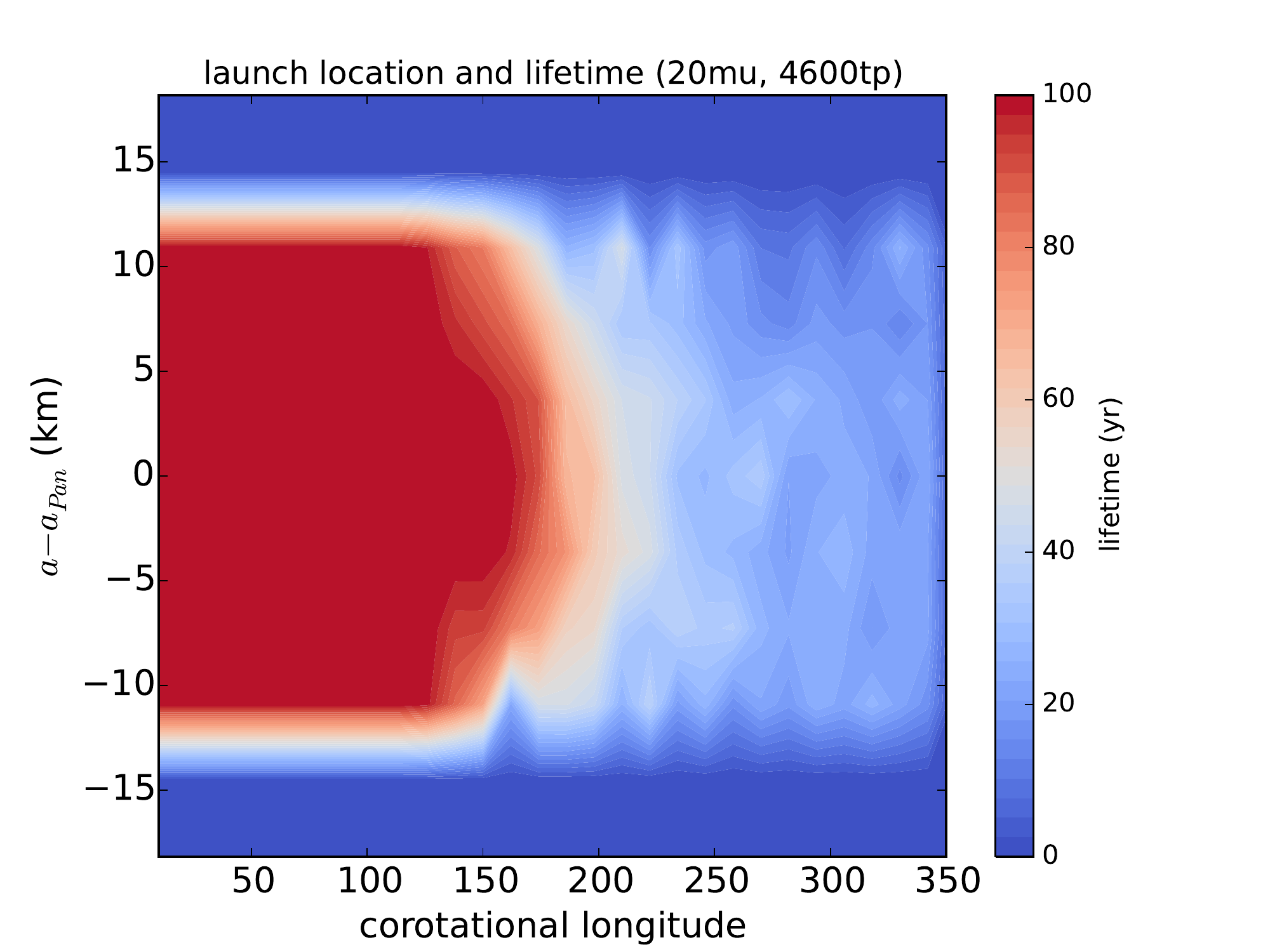}
        \end{subfigure}
        \begin{subfigure}
            \centering
            \includegraphics[width=0.48\textwidth]{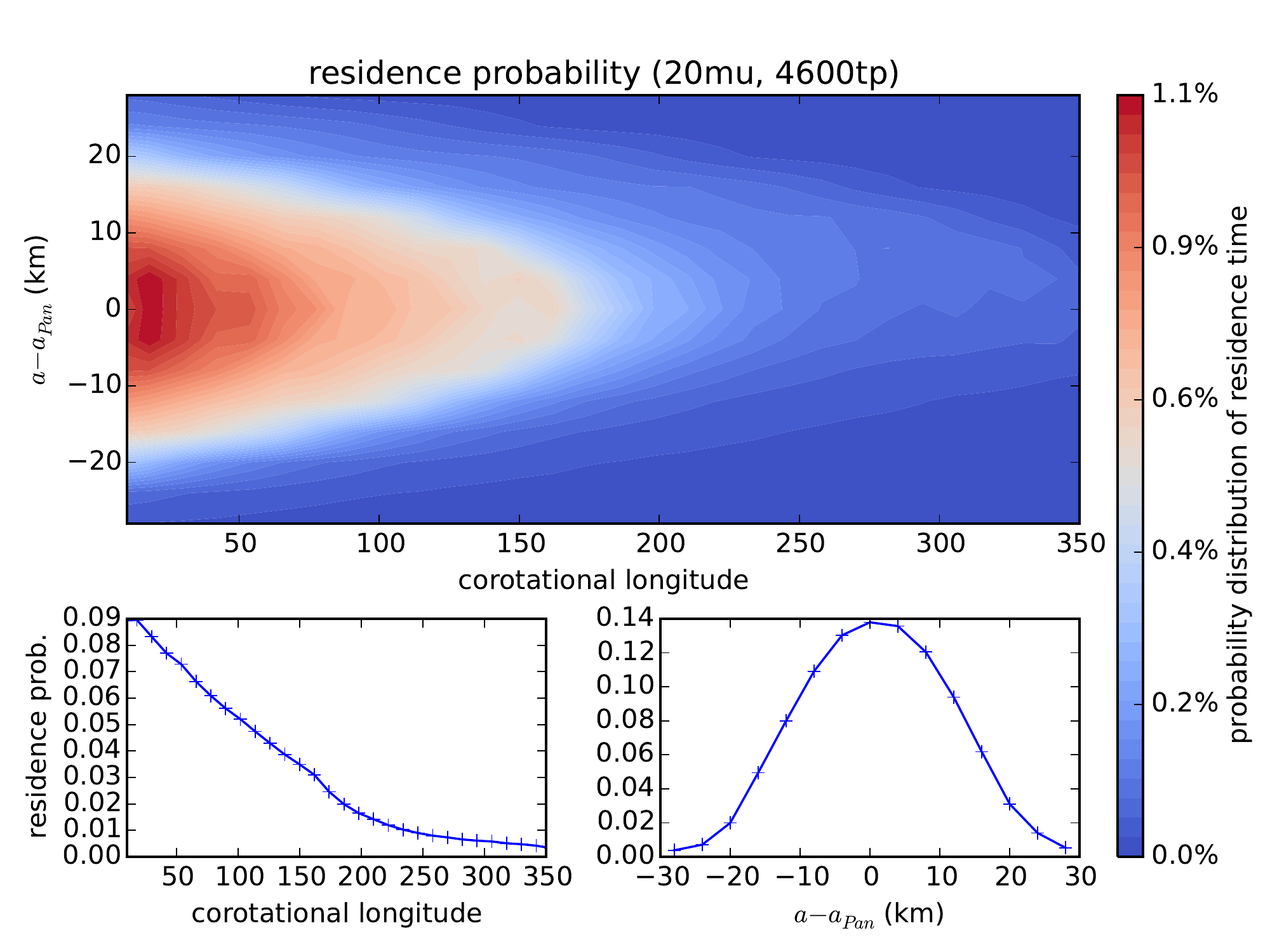}
        \end{subfigure}
        \caption{The restriction of plasma drag act on particle orbits, demonstrated by relation between particle lifetime and it's birth region (left panel) and the relative density of particles (right panel). There are two set of simulations: the upper panels are result of 10$\mum$ particles, the lower panels are for 20$\mum$. For each size there are 4,600 particles. All four figures are in corotational frame of Pan, and Pan is located at origin in these figures. The ion densities are same in four panels ($n_{\Opp} = 30 \cm^{-3}$). The grid size for contour plots are $\sim$12 deg $\times$ 4$\km$.}
        \label{fig:plasma_contour}
    \end{figure*}

    To quantify the restrictions the plasma drag exerts on ringlet particles, we simulate particle orbits of particles with size 10$\mum$ and 20$\mum$, for each size there are 4,600 particles. These particles originate randomly from  $\abs{a - a_{Pan}} < 10$\km, with small ($<10^{-6}$) initial eccentricities and inclinations.
    The simulation time is 100 years, which is longer than the average collisional lifetime of particles in the presence of our putative moonlet belt (here we ignore moonlets which act as sinks in these simulations).
    The particles are removed from simulation once their orbital radius is outside the Encke gap.
    The simulation result is shown in Fig.~\ref{fig:plasma_contour}.

    The left panels of Fig.~\ref{fig:plasma_contour} show the relationship between the birth place of particles and their lifetimes, the upper panel is for 10 micron radius particles, the lower one is for 20 micron particles.
    When the plasma drag is strong enough, particles generated from large corotational longitudes are not confined in the modified horseshoe orbit and are discarded in a very short time. This is because, once they leave the horseshoe region, they will eventually encounter Pan at larger semi-major axes and being gravitationally scattered to another orbit.
    On the other hand, particles from smaller corotational longitudes are restricted to a small region. The size of the region depends on the strength of drag force: the stronger the force, the smaller the longitudinal distribution.

    The normalized probability distribution of particle residence time is displayed in the right panels of Fig.~\ref{fig:plasma_contour}.
    It is obtained by counting the amount of particles in each grid for each time step and accumulate over the whole simulation time and then normalized.
    In these figures there are asymmetries in both azimuthal and radial direction.
    The azimuthal asymmetry is expected as the result of modified horseshoe orbit.
    The smaller asymmetry in radial direction is because of outward migration.

    In conclusion, the simulation result shows the combination of gravitational forces of Pan and Saturn with plasma drag cause a larger probability of finding ejected particles in a region leading Pan, as has been observed by Cassini and proposed by \citet{Hedman2013}.

\section{Conclusion and outlook}
    \label{sec:conclusion}
    The results of our investigations are summarized in Table~\ref{tbl:taus}.
    From this we conclude that impact-ejecta process with target of Pan and a putative embedded moonlet belt cannot sustain the Encke central ringlet, at least for the parameters we adopted here.
    The secondary dust sources \citep{DikarevKrivovEtAl2006} can enhance the particle production rate by almost one order of magnitude, but still not sufficient.
    Increase the number of embedded moonlets cannot increase the dust in the steady state ringlet, because moonlets act as dust sources and sinks simultaneously (see eq.~\eqref{eq:mpr} and \eqref{eq:T_moon}).
    Consequently, other mechanisms must be at work to replenish the Encke central ringlet.
    For example, moonlet mutual collisions may provide dust in the same order as the outputs of impact-ejecta model, although the amount of dust release per collision and the size distribution in the moonlet belt needs further investigation.

    \begin{table*}
        \centering
        \begin{tabular}{ccccl}
            \hline
            Source & $\dot{\tau}^+$ & $\angles{T_{life}}$ & $\tau$ & Note\\
            \hline
            Pan      & $3.6 \times 10^{-10}$   & 56.6   & $2.1 \times 10^{-8}$   & if $\eta = 0.125\%$ \\
            moonlets & $3.3 \times 10^{-7}$    & 56.6   & $1.9 \times 10^{-5}$   & \small{(1)} \\
            moonlets (collision) & $2.6 \times 10^{-8}$ & 56.6 & $1.5 \times 10^{-6}$ & \small{(2)} \\
            \hline
        \end{tabular}
      \caption{Summary of the optical depths at steady-state base on different models.
            The first two sources are about impact-ejecta process and the last one is related to moonlet mutual collision.
            The $\dot{\tau}^+$ is optical depth increase per year and the unit of $\angles{T_{life}}$ is years.
            The variable $\eta$ used in Table~\ref{tbl:n_plus} defines the fraction of new born dust that stay sufficient long time in the ringlet, the value adopt here is just a guess.
            Notes:
                   (1)The secondary source enhancement factor ($\sim 9.7$) is included.
                   (2)Assuming moonlets size ranged from 10--250 meter with $\gamma=3$ and the total cross section same as above. See the text for more details.
            }
      \label{tbl:taus}
    \end{table*}

    The particle dynamics can explain the minimum particle size and the azimuthal asymmetry in Encke central ringlet.
    The solar radiation pressure indicates that particle radius must be larger than 2.4$\mum$ to avoid collision with the gap edges.
    The plasma drag causes the semi-major axes of ringlet particles of 1--100$\mum$ increase by 10--0.1 km/year, and the uncertainties on ion densities may increase the number by a factor of 10--100.

    \citet{Hedman2013} proposed that the outward migration and the resulting azimuthal asymmetry of Encke central ringlet particles is caused by the combination of gravity of Saturn, Pan, and drag force (especially the plasma direct collision). Here we further consider the Coulomb plasma drag and use simulation to demonstrate the azimuthal asymmetry as the result of `modified horseshoe orbit'.
    Furthermore, we also shown that when the plasma drag is strong enough, the particle source is restricted to a small region of longitudes since particles generated from another region of the ringlet cannot survive for very long time --- they are pushed outside the horseshoe orbits.
    This implies one might deduce the strength of the plasma drag from the Cassini observations by studying the azimuthal asymmetry of the Encke central ringlet.

    In future work, we would like to further investigate the ring kinetics.
    A generalized concept of ring kinetics describes the balance of the counteracting processes of coagulation and fragmentation \citep{SpahnAlbersEtAl2004}.
    \citet{BodrovaSchmidtEtAl2012} have applied this concept to a case of bimodal size distribution (`carriers' and `dust') of ring particles, in order to explain the lack of sub-centimeter particles in dense planetary rings.
    However, this simplified model had not considered the shearing forces during lateral collisions and the direct actions by colliding carrier, both should increase the efficiency on releasing particles.

\section*{Acknowledgments}
    We thank to Miodrag Srem{\v c}evi{\'c} for the observation on the azimuthal variation in the Encke central ringlet.
    Also thank for Thomas M\"unch for the calculation in equilibrium potential.
    The work is supported by DFG SP 384/21-1.


\end{document}